%% file: paper.tex
\documentclass{elsarticle}
\usepackage[top=0.75in, bottom=1in, left=0.75in, right=0.75in]{geometry}
\usepackage[utf8]{inputenc}
\usepackage{todonotes}
\usepackage{makecell}
\usepackage{lscape}
\usepackage{xcolor}
\usepackage{tcolorbox}
\usepackage{enumitem}
\usepackage{afterpage}
\usepackage{amsmath}

\makeatletter
\renewcommand\paragraph{\@startsection{paragraph}{4}{\z@}%
	{3.25ex \@plus1ex \@minus.2ex}%
	{-1em}%
	{\normalfont\itshape}}
\makeatother

\newtcolorbox{findingsbox}{
	colback=gray!10,  % Light gray background
	colframe=black,   % Black border
	boxrule=1pt,      % Border thickness
	arc=5pt,          % Rounded corners
	left=10pt, right=10pt, top=5pt, bottom=5pt, % Padding
	width=0.95\textwidth, % Box width
}

\newcommand{\finding}[2]{
	\begin{findingsbox}
		\textbf{\underline{#1:}} #2
	\end{findingsbox}
}

\newcommand{\textapproxnew}{\raisebox{0.5ex}{\texttildelow}}

\journal{Information and Software Technology}

\begin{document}

\begin{frontmatter}

\title{Previously on... Automating Code Review\tnotemark[t1]}
\tnotetext[t1]{Preprint. Under review at \textit{Information and Software Technology} (Elsevier).}

\author{Robert Heumüller\tnotemark[t2]}
\ead{robert.heumueller@ovgu.de}
%\ead[url]{\href{https://orcid.org/1234-5678-9012}}
\tnotetext[t2]{Corresponding Author}

\author{Frank Ortmeier}
\ead{frank.ortmeier@ovgu.de}
%\ead[url]{\href{https://orcid.org/1234-5678-9012}}

\affiliation{organization={Otto von Guericke University},
             addressline={Chair of Systems and Software Engineering, Department of Computer Science,\\ Gustav-Adolf-Straße 15},
             city={Magdeburg},
             postcode={39106},
             state={Sachsen Anhalt},
             country={Germany}}

\input{content/abstract}

\end{frontmatter}

\input{content/introduction_new}
\input{content/questions_short}
\input{content/tasks}
\input{content/survey}
\input{content/results}
\input{content/experiment}

\input{content/threats}
\input{content/relatedwork}

\input{content/conclusion}

\section{Declaration of Generative AI and AI-Assisted Technologies in the Writing Process}
The authors used OpenAI ChatGPT-4o only for language proofreading and editing. 
All content was reviewed and edited, and the authors take full responsibility for the final manuscript.

\bibliographystyle{elsarticle-num-names} 
%\bibliography{literature,scopus/scopus}{}
\bibliography{cited}{}

\end{document}

%% file: content/abstract.tex
\begin{abstract}
\textbf{Context:}
Modern Code Review (MCR) is a standard practice in software engineering, yet it demands substantial time and resource investments.
Recent research has increasingly explored automating core review tasks using machine learning (ML) and deep learning (DL).
As a result, there is substantial variability in task definitions, datasets, and evaluation procedures.\\
\textbf{Objective:}
This study provides the first comprehensive analysis of MCR automation research, aiming to characterize the field’s evolution, formalize learning tasks, highlight methodological challenges, and offer actionable recommendations to guide future research.\\
\textbf{Methods:}
Focusing on the primary code review tasks, we systematically surveyed 691 publications and identified 24 relevant studies published between May 2015 and April 2024.
Each study was analyzed in terms of tasks, models, metrics, baselines, results, validity concerns, and artifact availability.
Additionally, we developed two simple baseline models for change quality estimation using synthetic features to provide a human-understandable reference for interpreting model performance.\\
\textbf{Results:}
We provide formal definitions of the primary code review tasks: change quality estimation, comment generation, and code refinement.
Our analysis reveals significant potential for standardization, including 48 task metric  combinations, 22 of which were unique to their original paper, and limited dataset reuse.
Aggregated results show that, while improvements were made, practical applicability remains limited.
We highlight challenges and derive concrete recommendations for examples such as the temporal bias threat, which are rarely addressed so far.
Additionally, 15 of the 24 approaches used only black box models as baselines.
Our experiment with synthetic feature baselines underscores the importance of including human understandable models, as these simple baselines matched or outperformed more complex ones and provide a more informative frame of reference.\\
\textbf{Conclusion:}
Our work contributes to a clearer overview of the field, supports the framing of new research, helps to avoid pitfalls, and promotes greater standardization in evaluation practices.
\end{abstract}

%% file: content/introduction_new.tex
\section{Introduction\label{sec-intro}}
Modern Code Review (MCR) is widely practiced in software development, from open source projects to companies like Microsoft and Google~\cite{10.5555/2486788.2486882,10.1145/3183519.3183525}.
However, studies show that it requires substantial time investment~\cite{Bosu2013}.
In response, AI4SE research has increasingly explored automating code review activities~\cite{Tufan2021}.
Although the field is still young and rapidly evolving, we present the first comprehensive study on the evolution and state of code review automation, covering 24 articles published between May 2015 and April 2024.

This study aims to provide a comprehensive analysis of MCR automation research to help researchers identify trends, define scopes, adopt methodologies, and avoid pitfalls.

We focus on the primary code review automation tasks, refining the terminology introduced by Z. Li et al.
These tasks include change quality estimation, comment generation, and comment implementation~\cite{Li2022}.
We consider them primary because they represent the activities where developers invest the most active time, offering high potential for efficiency gains through assistance or automation.
Other related tasks, such as reviewer recommendation~\cite{Cetin2021} or intelligent nudges~\cite{Maddila2023}, are not addressed here.
For a broader overview of MCR research, see Yang et al.~\cite{Yang2024}.

Key findings include (1) a highly diverse set of 48 task-metric combinations, many appearing only once, (2) limited reuse of standardized datasets and data collection methods, complicating performance assessment, and (3) a need for greater awareness of critical validity concerns like temporal bias and human-understandable baselines.

Our contributions are:
\begin{itemize}
    \item A systematic survey of 691 publications, identifying 24 relevant approaches since 2015, classified by a formalized definition of learning tasks and key contributions.
    \item An analysis answering four research questions aligned with our study objectives.
    \item Concrete recommendations on metric selection, result reporting, and addressing validity challenges.
    \item Two reusable baseline models for change quality estimation that, using trivial synthetic features, can match the performance of transformer baselines, emphasizing the importance of reference frames in result interpretation.
    \item Full publication of survey results and source code\footnote{We will upload to Zenodo upon acceptance. Reviewers can access the replication package on EasyChair} to enable replication and further research.
\end{itemize}

The remainder of the paper is structured as follows:
Section~\ref{sec-rq} states our four research questions.
Section~\ref{sec-tasks} formalizes the primary code review automation tasks.
Section~\ref{sec-survey} describes our survey method and results, answering the first research question.
Section~\ref{sec-discussion} answers the remaining three questions, highlights challenges, and offers corresponding recommendations.
Section~\ref{sec-experiment} introduces baseline models for change quality estimation and discusses their importance.
Section~\ref{sec-related-work} reviews related work, including other MCR surveys.
We conclude with threats to validity (Section~\ref{sec-threats}) and final remarks (Section~\ref{sec-conclusion}).

%% file: content/questions_short.tex
\section{Research Questions\label{sec-rq}}
To structure our investigation and ensure a focused analysis, we define the following four research questions and explain how they align with our study objective.
They guide our analysis by categorizing existing approaches, evaluating the use of metrics, assessing reported results, and identifying validity challenges.

\begin{enumerate}
	\item \emph{RQ1: Overview of MCR Tasks and Approaches} Which primary MCR tasks are the surveyed publications aimed at, and what concepts did they explore? 
	We formally define the primary MCR automation tasks in Section~\ref{sec-tasks} and use these definitions to categorize each surveyed approach.
	For each publication, we provide an overview of its concept, methodology, and results.
	Answering this question offers an overview of MCR automation research and provides definitions and a nomenclature for distinguishing different types of approaches.
	
	\item{RQ2: Metrics} Regarding the different MCR tasks, which metrics are used to measure quality of approaches? 
	Based on these findings, what recommendations can we derive?
	Answering this question consolidates information on metric usage, helps researchers select appropriate metrics, and advises on considerations when introducing new ones.
	
	\item{RQ3: Results} What do the reported results reveal about performance across tasks and metrics?
	We compile and organize quantitative results from all 24 surveyed publications, each possibly addressing multiple MCR tasks and using different datasets, to provide a structured view of the research landscape.
	Answering this question helps assess the overall state of performance and highlights challenges in comparing results across studies.
	
	\item{RQ4: Validity} Which validity challenges do the surveyed publications report, and what practices are suggested or can be derived to ensure research validity?
	Answering this question aggregates perspectives from multiple publications to derive concrete advice on potential pitfalls and avoidance strategies.
\end{enumerate}

%% file: content/tasks.tex
\section{Formalization of the Primary MCR Tasks\label{sec-tasks}}
\begin{figure}[h]
	\centering
	\includegraphics[width=0.8\linewidth]{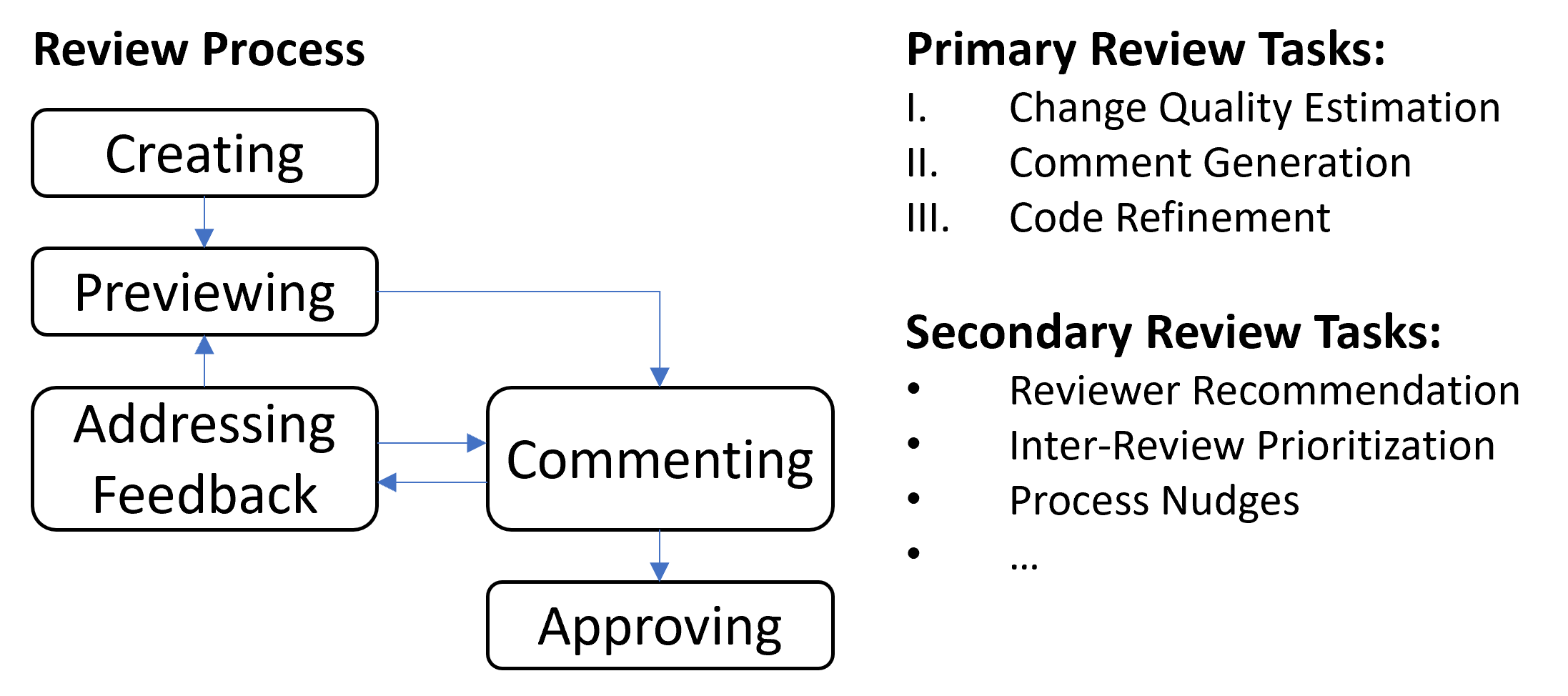}
	\caption{Code Review Process and Learning Tasks\label{fig-process-and-tasks}}
\end{figure}

To address our first research question, we formally define the primary code review automation tasks.
Figure~\ref{fig-process-and-tasks} shows the review process~\cite{10.1145/3183519.3183525} and the corresponding tasks, adopting the basic naming of Li et al.~\cite{Li2022}.
As outlined in the introduction, we focus on three primary tasks: Change Quality Estimation (ChQual), Comment Generation (ComGen), and Code Refinement (CodeRef).

We formalize the automation tasks with the following definitions:
\begin{align*}
	F = &\{f_1, ..., f_w\} &\text{Code Fragments}\\
	D = &\{d_1, ..., d_x\} = \{(f_i, f_j) | f_i != f_j\} &\text{Changes}\\
	R = &\{ r_1, ..., r_y | r_i \subseteq D\} &\text{Review Request}\\
	\mathit{C} = &\{\mathit{c_1},...,\mathit{c_z}\} &\text{Comment Phrases}\\
\end{align*}
A \emph{code fragment} \(f\) is a contiguous sequence of source code tokens.
The \emph{fragment granularity} indicates whether an approach works at token, line, method, class, or another level.
A \emph{change} \(d\) is a tuple of two fragments, original and changed, regardless of representation (e.g., diff-syntax or revisions).
Each \emph{review request} \(r\) consists of a set of changes submitted for review, covering pull-request-based and other processes.
Reviewers express reactions in \emph{comments} \(\mathit{c}\), considered atomic here, with vocabulary and syntax abstracted away.

Due to significant variability in how \emph{ChQual}, \emph{ComGen}, and \emph{CodeRef} are implemented in the surveyed literature, we further qualify tasks by \emph{fragment granularity}, review \emph{context} (F: Fragment, C: Change, R: Review), and possible \emph{focus} on parts of a change (T: Tokens, L: Lines).

Using this framework, we define the most common learning tasks and explain their application at different stages of the review process.

\paragraph*{Task I. Change Quality Estimation}
\begin{align*}
	\mathit{ChQualR}:\quad& \{(d,r) | d \in r \} \rightarrow  [0, 1] & \text{Review Context}\\
	\mathit{ChQualC}:\quad& D \rightarrow [0, 1] & \text{Change Context}\\
	\mathit{ChQualF}:\quad& F \rightarrow [0, 1] & \text{Fragment Context}
\end{align*}
The full task of \emph{ChQualR} is to decide whether a particular change \(d\) in the context of a code review \(r\) can be accepted or requires further work or discussion.
Formally, the objective is to learn a function \( \mathit{ChQual} \) that scores changes, indicating the certainty that they would be accepted without comments (0) or commented upon (1).
In our survey, we found that approaches considering only the change (\emph{ChQualC}) or fragment context (\emph{ChQualF}) are the most common.
In contrast, human reviewers consider a much broader set of features when deciding whether to accept or comment on a change, especially prior in-review discussions.
However, to the best of our knowledge, no existing approaches incorporate prior conversation.
Further, none of the surveyed approaches employed an additional line or token \emph{focus} for \emph{ChQual}.
Nonetheless, we can envision applications during the previewing and commenting phases, where querying a model at different granularities could assist human experts.

In the review process, automated change quality estimation models could first be used before the review request is published, either directly in the \emph{creating} phase (e.g., via an IDE plugin) or later during the \emph{previewing} phase in the review or pull-request tool.
This would allow contributors to identify and revise problematic code before submission.
Second, quality scores could help prioritize changes, enabling reviewers to focus on potentially critical code first.
As Fregnan et al. showed, the typical alphabetic order in which changed files are presented correlates significantly with the number of review comments they receive.
Likely causes include developers' limited and degrading attention during long reviews~\cite{Fregnan2022_1}.
Thus, change quality estimation models could support more efficient use of limited human resources.

\paragraph*{Task II. Review Comment Generation} 
\begin{align*}
	\mathit{ComGenR}:\quad& \{(d,r) | d \in r \} \rightarrow  \mathit{COM} & \text{Review Context}\\
	\mathit{ComGenC}:\quad& C \rightarrow \mathit{C} & \text{Change Context}\\
	\mathit{ComGenF}:\quad& F \rightarrow \mathit{C} & \text{Fragment Context}\\
\end{align*}

This task involves generating comment phrases \(\mathit{com}\) for a given change \(d\) in a review \(r\), emulating a human expert’s reaction—such as satisfaction, criticism, or suggestions.
In our survey, we found approaches only for \emph{ComGenC} and \emph{ComGenF}, beginning with Gupta et al. in 2018~\cite{Gupta2018}.
Future work could extend comment generation models to incorporate the full review context.

Regarding \emph{focus}, some approaches provide additional information on which particular lines should be commented on~\cite{Li2022a, Lu2023, Gupta2018}.
We refer to these as \emph{ComGenC+L} or \emph{ComGenF+L}.

Automated comment generation models can assist contributors in the creating and previewing phases and reviewers during the commenting phase.
Line-by-line analyses (\(\mathit{ComGenC+L}\) or \(\mathit{ComGenF+L}\)) could be especially useful, when combined with quality estimation models.

Many text generation models allow suggesting a set of the most likely comments, greatly improving the chance of producing a relevant one.
As Li et al. propose, this could eventually also let reviewers select a comment from suggestions rather than writing it themselves~\cite{Li2022}.
\paragraph*{Task III. Code Refinement}
\begin{align*}
	\mathit{CodeRef1R}:\quad& \{(d, \mathit{c}, r) | d \in r \} \rightarrow  F & \text{Review Context (With Comment)}\\
	\mathit{CodeRef1C}:\quad& D \times C \rightarrow  F & \text{Change Context (With Comment)}\\
	\mathit{CodeRef1F}:\quad& F \times C \rightarrow  F  & \text{Fragment Context (With Comment)}\\
	\\
	\mathit{CodeRef2R}:\quad& \{(d, r) | d \in r \} \rightarrow  F & \text{Review Context (Without Comment)}\\
	\mathit{CodeRef2C}:\quad&  D \rightarrow  F & \text{Change Context (Without Comment)}\\
	\mathit{CodeRef2F}:\quad&  F \rightarrow  F & \text{Fragment Context (Without Comment)}\\
\end{align*}	

This task involves, given a change \( d=(f_1,f_2) \) in a review \(r\) and a particular comment \(\mathit{c}\), generating a refined version \(f_3\) that satisfies the intention expressed by the comment as precisely as possible.

The first task explored was \(\mathit{CodeRef2F}\), directly learning the transformation from original to revised code without using review comments, as studied by M. Tufano et al.~\cite{Tufano2019}.
Intuitively, this approach is limited, since starting at some revision, many different changes are legitimate and possible.
More recently, researchers in the same group showed that informing the model with comments (\(\mathit{CodeRef1}\)) improves performance compared to \(\mathit{CodeRef2}\)~\cite{Tufan2021, Tufano20222291}.

Regarding \emph{focus}, some approaches used additional information on the particular subset of lines (\emph{CodeRef1F+L}) or tokens (\emph{CodeRef1F+T}) that were subject to the review comment~\cite{Lu2023, Tufano20222291, Tufan2021}.

\emph{CodeRef} models can be applied at various stages of the review process, primarily assisting contributors either by suggesting improvements before submission or by implementing reviewer-suggested changes (as proposed by~\cite{Li2022}).
Additionally, examining implementation candidates could benefit reviewers by illustrating the kinds of code changes their comments might trigger.

%% file: content/survey.tex
\section{Survey\label{sec-survey}}
Having formalized the primary code review tasks in Section~\ref{sec-tasks}, we now address our first research question through a systematic literature survey. 
We begin by outlining our methodology in Section~\ref{sec-method}, followed by an overview of each individual publication. 
Our focus is on the most relevant aspects of each paper concerning our research questions and the broader context, recognizing that not every detail can be covered.
The remainder of this section is structured by \emph{non-transformer models} (Section~\ref{sec-methods-non-transformer}), \emph{transformer models} (Section~\ref{sec-methods-transformer}), and \emph{retrieval-based methods} (Section~\ref{sec-methods-retrieval}).
Transformer and non-transformer refer to the primary model architecture, while \emph{retrieval methods} are a distinct group that recommend review comments or code changes from historical review data, without adapting them to the new context.

Table~\ref{fig-survey-overview} presents an overview of all surveyed articles, including their learning tasks, fragment granularity, model architecture, and availability of research artifacts.
Complementing Table~\ref{fig-survey-overview}, we also provide a supplementary table with additional information including availability details, datasets, baselines, results and reported validity considerations alongside this paper.
This is the raw data for the discussion of RQ2-RQ4 in Section~\ref{sec-discussion}.

\afterpage{
\begin{landscape}	
	\begin{table}
		\resizebox{0.88\linewidth}{!}{
			\begin{tabular}{llp{2cm}p{1cm}p{1cm}p{1cm}p{1cm}lp{4cm}ll}
				Paper                                    & Year & Name                   & Ch\-Qual & Com\-Gen & Code\-Ref1 & Code\-Ref2 & \makecell[l]{Fragment\\Granularity} & Architecture                         &\makecell[l]{Availability} \\
				\hline
				~\cite{Gupta2018}        & 2018 & DCR &        & F+L    &          &          & Lines                & LSTM                                 &              \\
				~\cite{Guo2019}          & 2019 & Rsharer                &        & F      &          &          & Lines                & CNN                                  & D            \\
				~\cite{Li2019}           & 2019 & DeepReview             & C,R    &        &          &          & Lines                & CNN                                  &              \\
				~\cite{Shi2019}          & 2019 & DACE                   & C+L    &        &          &          & Lines                & CNN+LSTM                             &              \\
				~\cite{Tufano2019}       & 2019 & Tufano2019             &        &        &          & F        & Method               & RNN                                  & M+C+D        \\
				~\cite{Guo2020}          & 2020 & Rsharer+               &        & F      &          &          & Lines                & CNN                                  &              \\
				~\cite{Siow2020}         & 2020 & CORE                   &        & C      &          &          & Lines                & LSTM                       &              \\
				~\cite{Staron2020513}       & 2020 & Auto\-Code\-Reviewer       & F      &        &          &          & Lines                & Random\-Forest, CNN+LSTM   & C+D          \\
				~\cite{Hellendoorn2021}  & 2021 & Hellendoorn\-2021        & R      &        &          &          & Lines                & Transformer                          & C            \\
				~\cite{Tufan2021}       & 2021 & Tufano2021             &        &        & F+T      & F        & Method               & Transfomer (OpenNmt)                 & M            \\
				~\cite{Hong20221034}      & 2022 & RevSpot                & F      &        &          &          & Lines                & Random\-Forest, LIME                   & C            \\
				~\cite{hong2022commentfinder}      & 2022 & CommentFinder          &        & F      &          &          & Method               & Retrieval (BoW + KNN)      & C            \\
				~\cite{Li2022}          & 2022 & GCTC                   &        &        &          & F        & Method               & Transformer                          & \emph{broken link}  \\
				~\cite{Li20221035}          & 2022 & CodeReviewer           & C      & C      & C        &          & Lines                & Transfomer (T5)                      & M+C+D        \\
				~\cite{Li2022a}          & 2022 & AUGER                  &        & F+L    &          &          & Method               & Transformer (T5)                     & C            \\
				~\cite{Thongtanunam2022} & 2022 & Auto\-Transform          &        &        &          & F        & Method               & Transformer                          & C+D          \\
				~\cite{Tufano20222291}       & 2022 & Tufano2022             &        & F      & F+L      & F        & Method               & Transformer (T5)                     & M+C+D        \\
				~\cite{Wu2022}           & 2022 & SimAST-GCN             & C      &        &          &          & Method               & GCN                                  & C+D          \\
				~\cite{Lin2023b}        & 2023 & LinHY2023              &        &        &          & F        & Method               & \makecell[l]{Transformer\\(Code-T5, GraphCodeBERT)} &              \\
				~\cite{Lu2023}           & 2023 & LLaMA\-Reviewer         & C      & F+L, C & F+L, C   &          & Lines                & LLM (LLaMA 6.7B)                     & M+C+D        \\
				~\cite{Shuvo2023}        & 2023 & RevCom                 &        & C      &          &          & Lines                & Retrieval (TF-IDF + BM25)              & D            \\
				~\cite{Yin2023}          & 2023 & crBERT                 &        &        & F        & F        & Method               & Transformer (Code\-BERT)               & D            \\
				~\cite{Zhou2023}         & 2023 & Zhou2023               &        & F      & F+L      & F        & Method               & Transforme (CodeT5)                  & C+D          \\
				~\cite{Cao2024}          & 2024 & SMILER                 &        &        & F        & F        & Method               & Transformer                          & C+D         
		\end{tabular}}
		\caption{Overview of surveyed publications. Availability specified for Model (M), Code (C), and Data (D).\label{fig-survey-overview}}
	\end{table}
\end{landscape}}

\subsection{Method\label{sec-method}}
\begin{figure*}
	\includegraphics[width=\linewidth]{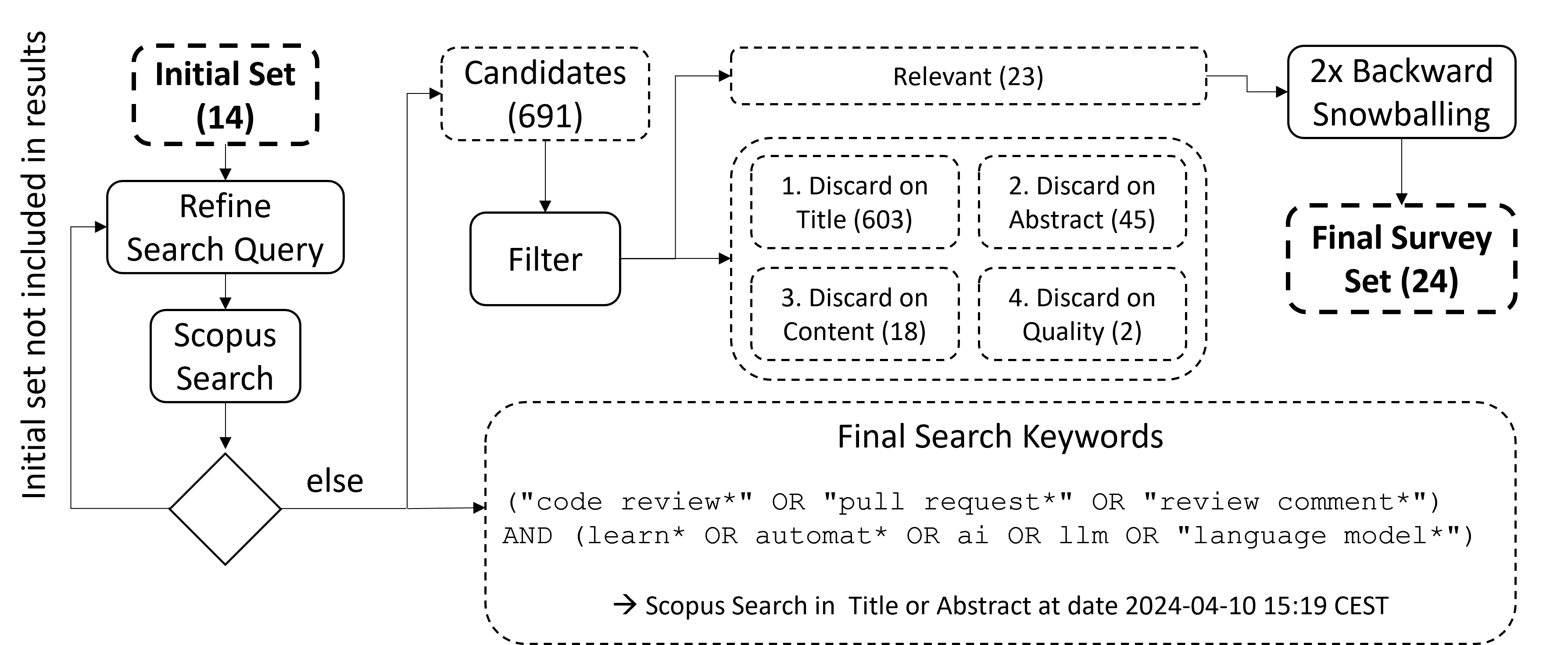}
	\caption{Systematic Literature Review Method\label{fig-survey-slr}}
\end{figure*}

To identify relevant publications for the survey, we followed a process based on the guidelines by Kitchenham et al.~\cite{KITCHENHAM20132049}.
First, we defined our inclusion criterion as: \emph{Publications reporting quantitative results on models automating one of the primary MCR tasks described in Section~\ref{sec-tasks}}.
This also excludes approaches generating entirely synthetic comments based on static analysis, such as Balachandran's study~\cite{Balachandran2013}.
Second, starting from an initial set of 14 known relevant publications from our prior work, we constructed a search query for \emph{Scopus}\footnote{www.scopus.com}, targeting titles and abstracts.
We incrementally added and abstracted keywords using wildcards until all known publications were included.
The final query, shown in Figure~\ref{fig-survey-slr}, was executed on the afternoon of April 10, 2024.
Starting with 691 candidate publications, we applied four filtering steps based on title, abstract, content, and, in two cases, paper quality.
From this, we identified 23 publications, followed by two iterations of manual backward snowballing, which added one more matching publication.

Each of the 24 final publications was then analyzed with regard to the four research questions defined in Section~\ref{sec-rq}. 
Specifically, for each publication, we classified the learning tasks addressed, the fragment granularity, and line- or token-focus of the study and further recorded:

\begin{itemize}
	\item The name of the approach or model
	\item The model type, architecture, and whether and availability information
	\item Details on data collection and pre-processing procedures, in particular filtering, augmentation and balancing
	\item Dataset size and availability, including training/validation/testing splits
	\item Baseline models used for comparison
	\item Evaluation metrics employed
	\item Reported results per learning task, metric, and dataset in a machine-readable way
	\item Notable statements regarding internal or external validity, significance testing, sources of bias, or temporal bias
	\item A concise summary of each publication's contribution and findings
\end{itemize}

Through this structured elicitation we achieve the necessary comparability across studies and provide the data for the analyses in the following sections.
All data is included in our replication package.

\subsection{Non-transformer Models\label{sec-methods-non-transformer}}
\noindent Before the widespread adoption of transformers, various architectures, such as CNN and RNN flavors, were the mainstay in text learning.
This subsection introduces MCR automation approaches using one of these types of models.

\paragraph*{DACE: Shi et al.~\cite{Shi2019}}
address the \emph{ChQualC+L} task, i.e. change quality estimation with change level context and additional line focus, at the lines-level fragment granularity.
It encodes code using CNN and LSTM layers at the statement and change levels, with a pairwise autoencoder (PAE)\footnote{Classical autoencoder, but it does not learn a sequence-to-sequence but rather two paired sequences to two paired sequences} extracting features from old and new code vectors.
The dataset comprises 35,640 hunks from six Apache Code Review Board projects, with strong imbalance (only 769 rejected hunks).
Baselines include a TFIDF-SVM and adaptations of \emph{Deeper}~\cite{Yang2015}.
Using 10-fold cross-validation\footnotemark[1], \emph{DACE} achieved 48\% \emph{F1} and 83\% \emph{AUC}, outperforming its baselines.

\paragraph*{DeepReview: H. Li et al. 2019~\cite{Li2019}}
address the \emph{ChQualR} task, i.e. change quality estimation with review level context, at line level fragment granularity and, unlike \emph{DACE}, uses a multi instance learning setup.
It encodes each hunk individually from diff syntax (old lines, new lines, and description), embeds them via CNNs, and merges their representations into a unified feature vector for classification through fully connected layers.
The dataset extends that of \emph{DACE} with a sixth Apache project, totaling \textapproxnew 29k accepted vs. \textapproxnew 630 rejected hunks.
Baselines include a TFIDF-SVM and modified versions of \emph{Deeper}~\cite{Yang2015}.
A 10-fold cross-validation\footnotemark[1] showed \emph{DeepReview} achieving 45\% \emph{F1} and 78\% \emph{AUC}, outperforming baselines; an ablation study confirmed benefits from multi-hunk classification.

\footnotetext[1]{No details were provided on how potential temporal biases from cross-validation were addressed. Temporal bias is further discussed in Section \ref{sec-discussion-validity}.}

\paragraph*{M. Tufano et al. 2019~\cite{Tufano2019}}
were to the best of our knowledge, the first to explore RNN models for automating changes between submitted and accepted versions in code review.
Precisely, they addressed \emph{CodeRef2F} at method-level fragment granularity.
The work evaluates LSTM and GRU models (no reported performance difference) on \textapproxnew 240k tuples of \emph{submitted\_method\_code} and \emph{accepted\_method\_code} originating from three large \emph{Gerrit} projects, filtered to \textapproxnew 22k instances by removing fragments over 100 tokens.
To reduce vocabulary size, identifiers and literals were replaced with generic names per instance, but preserving the 300 most frequent per project as idiomatic.
Evaluation used an 80/10/10 split and introduced the \emph{ExactMatches (EM)} metric (here called \emph{PerfectPredictions}), with TOP-1 to TOP-10 \emph{EM} ranging from 16--36\%.
A qualitative analysis of 722 exact matches identified learned patterns in refactorings, bug fixes, and common operations.

Despite limited input capacity and heavy preprocessing, this work introduced methods and evaluation practices that shaped later research.
Further, these early results already underscored challenges in interpreting performance scores given the lack of human-understandable baselines and data skew from heavy preprocessing.

\paragraph*{Auto\_code\_reviewer: Staron et al. 2020~\cite{Staron2020513}}
address the \emph{ChQualF} task, i.e. classifying whether individual lines need human review comments, at line-level granularity.
It assumes lines should be reviewed if they resemble lines that previously received negatively toned comments.
The dataset was extracted from Gerrit reviews across two open-source and two closed-source projects.
While the full dataset size is not reported, the available \emph{Wireshark} subset contains only 439 code-comment pairs, split 60\%/40\% for training and testing.
In preprocessing, each comment is broadcast to every line in the change, and a keyword-based sentiment detector assigns positive or negative labels.
Due to class imbalance (18\% positives in Wireshark), the training set was balanced via random upsampling.
They evaluated Adaboost and CNN models, with CNN achieving better results: \emph{MCC} of 0.7 and \emph{F1} of 0.95.
No baselines were used for comparison.

\paragraph*{RevSpot: Hong et al. 2022~\cite{Hong2022}}
address the \emph{ChQualF} task, i.e. predicting which lines in a changeset will receive comments or revisions, at line-level granularity.
It aims to reduce contributor wait times by providing early feedback.
An initial analysis of code review data showed median wait times of 15–64 hours across \emph{OpenstackNova}, \emph{OpenstackIronic}, and \emph{QtBase}, attributed to differences in patch size.
\emph{RevSpot} processes code fragments using bag-of-words features at both file and line levels.
File-level classification uses models such as random forests and decision trees.
For predicted positives, LIME~\cite{Ribeiro2016} highlights influential lines.
Data preparation includes SMOTE~\cite{10.5555/1622407.1622416} augmentation, and splits are based on historical order to avoid time bias.
File-level prediction achieved \textapproxnew~70\%–80\% accuracy.
At the line level, \emph{RevSpot} outperformed an n-gram baseline with 81\% top-10 accuracy, exceeding it by 56\%.

\subsection{Transformer Models\label{sec-methods-transformer}}
\noindent Transformer-based models have been the state-of-the-art in text learning for several years ~\cite{Vaswani2017}, and thus it is not surprising that the majority of recent approaches fall into this category.

\paragraph*{R. Tufano et al 2021~\cite{Tufan2021}}
Building on M. Tufano et al. ~\cite{Tufano2019}, this work continues the group's MCR automation research, following similar data preparation and evaluation and combining quantitative and qualitative methods.
It introduces two key changes: switching to a transformer model and analyzing the impact of including comments.
Special tokens mark the start and end of commented code, framing the task as \emph{CodeRef1F+T} at the method level.
Experiments used triplet data \((\mathit{submitted\_method\_code}, \mathit{comment}, \mathit{accepted\_method\_code}) \) from Gerrit and GitHub pull requests.
Preprocessing, adapted from the 2019 study, reduced over 150k entries to under 18k.
They trained a baseline \emph{CodeRef2F} model (without comments) and the new \emph{CodeRef1F+T} model.
The architecture was a dual-encoder transformer using OpenNMT ~\cite{Klein2018}.
Metrics included \emph{ExactMatches}, \emph{BLEU-4}~\cite{papineni2002bleu}, and normalized Levenshtein distance at the token level.
TOP-1 results showed strong gains from using comments: \emph{EM} increased from 3\% to 12\%, \emph{BLEU-4} from 77\% to 81\%, and normalized \emph{LEV} dropped from 24\% to 18\%.
The qualitative evaluation (not detailed here) again analyzed learned change types and the role of comments in generating better outputs.l as the types of comments that were helpful in generating superior code fragments.

\paragraph*{Hellendoorn et al. 2021~\cite{Hellendoorn2021}}
critically observed that the high degree of filtering, processing and re-balancing of earlier approaches like ~\cite{Tufano2019,Tufan2021} strongly limited the generalizability of their results for practical application. 
To better assess real-world challenges in automating code review, they proposed a new data collection strategy to build a larger triplet dataset (similar to ~\cite{Tufan2021}) with \textapproxnew 590k comment threads from 245 GitHub projects.
To reduce vocabulary size, they used \emph{sub-word splitting with byte-pair encoding} ~\cite{Karampatsis2020}, fixing the vocabulary at 25k tokens.
To avoid time biases ~\cite{Jimenez2019}, data was split both historically—ensuring all training instances precede testing—and across distinct project sets.
No final instance count was reported, and the dataset is not publicly available.

Their experiment studies \emph{ChQualR} at the line-level granularity, arguing this task is a prerequisite for others.
Without it, \emph{ComGen} and \emph{CodeRef} models assume prior knowledge of which changes will be commented on.
They apply a large transformer-style model that jointly processes all diff-hunks in a pull request to predict the probability of each fragment receiving a comment.
These probabilities are used to rank fragments during evaluation, considering both the trade-off between precision and recall at varying thresholds and the model's ability to rank all commented fragments above non-commented ones.
The model clearly outperforms a random baseline, but even the best configurations fail to reach 50\% overall precision, even at low recall.
They conclude that with ambiguous, highly imbalanced data\footnote{Only 10\% of code fragments in their dataset are commented.}, model performance likely cannot be meaningfully improved by scaling model size or increasing training time.
Instead, they recommend research focus on \emph{quantifying ambiguity} and rethinking \emph{modeling priorities}, especially in terms of what automated support developers truly need and whether simpler tasks could better deliver it.

\paragraph*{R. Tufano et al. 2022~\cite{Tufano20222291}}
more recently used a new dataset with less rigorous filtering, resulting in \textapproxnew 178k triplets from \textapproxnew 383k raw instances.
The main focus was on analyzing the effect of unsupervised pre-training for a T5-based transformer using \emph{masked language modeling}.
They created a separate pre-training dataset combining source code and technical English from CodeSearchNet ~\cite{Husain2019} and ~\cite{stackexchangedump}.
In addition to prior tasks, they also studied the \emph{ComGenF} task at the method level.
For evaluation, they replaced \emph{BLEU} with \emph{Code-BLEU}, a metric designed to better reflect code similarity.
Pre-training significantly improved performance on \emph{ComGenF} and \emph{CodeRef1F+L}, while T5 without pre-training performed better on \emph{CodeRef2F}.
They attribute this to \emph{CodeRef2F} not using comments, limiting benefits from joint pre-training on code and technical English.
As a baseline, the new T5 model was compared to their previous transformer ~\cite{Tufan2021}, which it generally outperformed.
Still, the results at TOP-1 on the large dataset remain too low for practical use: \emph{ComGenF} (2\% \emph{EM}, 8\% \emph{BLEU-4}), \emph{CodeRef1F+L} (14\% \emph{EM}, 81\% \emph{CodeBLEU}), and \emph{CodeRef2F} (5\% \emph{ExactMatches}, 82\% \emph{CodeBLEU}).
They emphasize that \emph{CodeBLEU} scores, while numerically high, are hard to interpret and mainly useful for comparing models relatively.

\paragraph*{AUGER: L. Li et al. 2022~\cite{Li2022a}}
address the \emph{ComGenF+L} task, i.e. generating review comments for specific lines in a changed method, at method-level granularity with explicit line marking.
It uses a T5-based transformer pre-trained on mixed Java code and review comments~\cite{Elnaggar2021}, extending ideas from R. Tufano et al.~\cite{Tufano20222291}.
Input consists of changed method code with special tokens marking lines to be commented.
The dataset, reduced from \textapproxnew46k to \textapproxnew10k instances, was prepared using identifier splitting based on naming conventions and a ninefold training augmentation via comment rewriting~\cite{Wei2019}.
Data was split 80\%/10\%/10\%.
AUGER outperformed both the R. Tufano et al. 2022 model~\cite{Tufano20222291} and a fine-tuned CodeBERT~\cite{Feng2020} on \emph{EM} and \emph{ROUGE-L}.
An ablation showed that removing line indicators or pre-training degraded performance, with line markers having the stronger impact.
A manual evaluation of 100 non-matching comments using a Bosu et al.~\cite{7180075} heuristic found 51\% of AUGER’s outputs useful, versus 62\% for human-written ones.

\paragraph*{CodeReviewer: Z. Li et al. 2022~\cite{Li2022}}
addresses the \emph{ChQualC}, \emph{ComGenC}, and \emph{CodeRef1C} tasks using a \emph{CodeBERT}-based model~\cite{Feng2020} trained to interpret diff syntax through unsupervised pre-training.
Thus the fragment granularity is diff level.
A key pre-training task is \emph{diff tag prediction}, where masked diff operations (ADD, DEL, KEEP) must be recovered.
Evaluation was performed on a new GitHub dataset, comparable in size to that of R. Tufano 2022~\cite{Tufano20222291}, including \textapproxnew 328k labeled, balanced\footnote{Balancing was performed using random down-sampling of non-commented hunks.} instances for change quality estimation.
Baselines included fine-tuned versions of R. Tufano 2022 and Code-T5~\cite{Tufano20222291,Wang2021}.
On \emph{ChQualC}, \emph{CodeReviewer} outperformed Code-T5 by 8.24\% \emph{F1} and 7.07\% \emph{ACC}.
For \emph{ComGenC} and \emph{CodeRef1C}, a human evaluation was conducted, with professionals rating top comment samples.
Results show that pre-training on diff syntax improves performance over generic transformers.
An ablation study confirmed the individual contribution of each pre-training task.

\underline{Remarks:} Absolute scores for \emph{ChQualC} are also reported and interpreted in Section~\ref{sec-experiment}.

\paragraph*{AutoTransform: Thongtanunam et al. 2022~\cite{Thongtanunam2022}}
address the \emph{CodeRef2F} task at method level, with a focus on handling newly introduced tokens.
To this end, it applies Byte-Pair Encoding (BPE)~\cite{Karampatsis2020} for subword tokenization, enabling generalization to unseen tokens via learned merge operations.
The resulting subword sequence is passed to a transformer-based neural machine translation model, which predicts target subwords using beam search.
Evaluation used the same dataset as M. Tufano et al.~\cite{Tufano2019}, distinguishing between changes without new tokens (2.508 instances) and with new tokens (12.242 instances).
Using the \emph{EM} metric, top-1 predictions achieved 14\% for methods without new tokens and 9\% for methods with new tokens, surpassing the Tufano baseline of 8\% and 0\%, respectively, highlighting the effectiveness of the central idea.

\paragraph*{SimAST-GCN: Wu et al. 2022~\cite{Wu2022}}
addresses the \emph{ChQualC} task at method level, predicting whether a change will be merged rather than whether it will receive comments.
It introduces an AST-based GCN model that simplifies ASTs by removing redundant nodes, uses a bi-GRU RNN for initial node embeddings, and applies a 3-layer GCN with attention pooling to generate method representations.
Each change is represented by subtracting the embeddings of the original and modified ASTs.
Evaluation used a custom dataset of 9 GitHub projects, with \textapproxnew 110k instances labeled with binary merge outcomes.
A 60\%/20\%/20\% split was used for training, validation, and testing.
Sub-datasets were highly imbalanced, with negative instance rates ranging from 6\% to 91\%.
The model achieved 81.21\% \emph{ACC} and 88.2\% \emph{F1}, outperforming all baselines (DACE, TBRNN, ASTNN, CodeBERT, GraphCodeBERT) by over 3\%.

\paragraph*{GCTC: Li et al. 2022~\cite{Li2022}}
addresses the \emph{CodeRef2F} task at method-level fragment granularity.
The model improves on Tufano’s RNN approach~\cite{Tufano2019} by addressing limitations in long-sequence handling and structural representation.
It introduces BNCS (Bi-directional Natural Code Structure), a graph format extending ASTs with token-order edges.
BNCS graphs are linearized via the SBT method~\cite{Hu2019} and processed with an encoder-decoder transformer.
The dataset includes method pairs from Gerrit’s Android, Ovirt, and Google projects, totaling 10,738 small and 10,991 medium-sized methods.
An implied 90\%/10\% train-test split was used.
Evaluation with \emph{EM} shows scores between 18\% (top-1) and 22\% (top-10) for medium methods across datasets.
The model outperforms the Tufano 2019 baseline~\cite{Tufano2019}, with gains ranging from 109\% to 343\%.
A qualitative analysis examined which change types were better handled than by the baseline.

\paragraph*{crBERT: Yin et al. 2023~\cite{Yin2023}}
focus on the \emph{CodeRef2F} and \emph{CodeRef1F} tasks at method-level granularity.
Their contribution lies in integrating structural information from Program Dependence Graphs (PDGs) with code and comments to improve transformer-based code generation.
PDGs are serialized and concatenated with source code and review comments, forming the input for fine-tuning a standard CodeBERT~\cite{Feng2020} model.
The study uses R. Tufano’s 2021 dataset and model as baseline~\cite{Tufan2021}.
Evaluation uses \emph{BLEU-4}, \emph{LEV}, and \emph{ROUGE-L} metrics and showed that \emph{crBERT} improved beyond the Tufano baseline.
For \emph{CodeRef1F}, the model achieves 78\% \emph{BLEU}, 93\% \emph{ROUGE-L}, and 18\% \emph{LEV}.
For \emph{CodeRef2F}, BLEU is not reported\footnote{There seems to be a copy-and-paste error in the paper.}, but scores include 94\% \emph{ROUGE-L} and 17\% \emph{LEV}.

\paragraph*{LLaMA-Reviewer: Lu et al. 2023~\cite{Lu2023}} 
were the first to apply a general-purpose large language model, LLaMA, to automate various code review tasks.
Their model targets \emph{ChQualC}, \emph{ComGenF+L}, \emph{ComGenC}, \emph{CodeRef1F+L}, and \emph{CodeRef1C} at line or method granularity, depending on the dataset.
The architecture uses a LLaMA transformer with 6.7 billion parameters.
It is first pre-trained on the CodeAlpaca dataset, then fine-tuned using parameter-efficient methods (PEFT), specifically LoRA and prefix tuning~\cite{Hu2021,Li2021}.

LLaMA-Reviewer is evaluated on the Tufano 2022 ~\cite{Tufano20222291} and CodeReviewer ~\cite{Li20221035} datasets, using several baselines including Tufano’s 2022 models, CodeReviewer, CodeT5, and CommentFinder ~\cite{Tufano20222291, Li20221035, Wang2021t5, hong2022commentfinder}.
Metrics include \emph{PRE}, \emph{REC}, and \emph{F1} for \emph{ChQualC}, and \emph{BLEU-4} for \emph{ComGenC} and \emph{CodeRef1C}.

On \emph{ChQualC}, the model achieves 60.99\% \emph{PRE}, 83.5\% \emph{REC}, and 70.49\% \emph{F1}, matching CodeReviewer on its original dataset ~\cite{Li20221035} while emphasizing recall over precision.

For \emph{ComGenC}, it scores 5.7\% \emph{BLEU-4} on the CodeReviewer dataset, outperforming all baselines, and 5.04\% on the Tufano dataset, where CodeReviewer and Tufano perform better.
This drop is attributed to formatting differences between the Tufano dataset and CodeAlpaca pre-training data.

On \emph{CodeRef1C}, LLaMA Reviewer reaches 82.27\% \emph{BLEU} on the CodeReviewer dataset and 78.23\% on Tufano, closely matching top baselines.

The study found that LoRA consistently outperforms prefix tuning, likely due to its greater number of trainable parameters.
Consistent with this assumption, the researchers observed performance improving with higher LoRA ranks.
Adding a language label to prompts gives a small but statistically significant gain in \emph{BLEU-4} (81.59\% vs. 82\%).

\paragraph*{Lin and Thongtanunam (2023)~\cite{Lin2023b}}
study whether structure-aware transformer models outperform standard ones at method-level \emph{CodeRef2F}.
The study compares CodeT5~\cite{Wang2021t5}, GraphCodeBERT~\cite{Guo2020a}, and Tufano’s 2022 T5 model~\cite{Tufano20222291} on datasets from ~\cite{Tufano20222291} and ~\cite{Thongtanunam2022}, totaling 332k instances with an 80\%/10\%/10\% train-validation-test split.
To enable generation, GraphCodeBERT’s encoder-only architecture was extended with a decoder.
All models were fine-tuned for method-level code refinement.
Metrics include \emph{EM}, \emph{Mean Reciprocal Rank (MRR)}, and \emph{Dataflow Matches (DFM)}.
At TOP-10, CodeT5 achieved 22\% \emph{EM}, GraphCodeBERT 18\%, and Tufano 10\%.
GraphCodeBERT led in \emph{MRR}, and both structure-aware models outperformed the baseline overall.
In \emph{DFM} at TOP-10, CodeT5 reached 33\%, GraphCodeBERT 30\%, and Tufano 22\%.
These results support an assumed advantage of structure-aware models for MCR automation.

\paragraph*{Zhou et al. (2023)~\cite{Zhou2023}}
investigate the performance of general purpose transformers pre-trained on code-related tasks for \emph{ComGenF}, \emph{CodeRef1F+L}, and \emph{CodeRef2F}, all at method level granularity.
They compare three task specific baselines, Tufano2021~\cite{Tufan2021}, Tufano2022~\cite{Tufano20222291}, and AutoTransform~\cite{Thongtanunam2022}, with general purpose models: CodeBERT~\cite{Feng2020}, GraphCodeBERT~\cite{Guo2020a}, and CodeT5~\cite{Wang2021t5}.
All models are fine-tuned on the original datasets used by the baselines.
To enable code generation, a decoder was added to the encoder only architectures of CodeBERT and GraphCodeBERT, following the structure of T5.

A key contribution of the study is the \emph{EditProgress (EP)} metric for \emph{CodeRef} tasks, complementing \emph{ExactMatches}.
\emph{EP} is based on edit distance and quantifies how much closer a model’s output gets to the ground truth, rather than measuring only full correctness.
The idea is illustrated as follows~\cite{Zhou2023}:

\begin{enumerate}
	\item Assume a submitted code fragment, \(f_{s}\), requires  \(k_1=4\) edit steps to transform it into the ground truth fragment, \(f_{gt}\).
	\item Now, a \emph{CodeRef} model generates a changed candidate fragment \(f_{c}\) that requires only \(k_2=3\) edit steps to transform it into  \(f_{gt}\).
	\item Then, the model made an \emph{EditProgress} of  \(\mathit{EP}=(k_1-k_2)/k_1=(4-3)/4=25\%\).
\end{enumerate}

The authors argue that \emph{EP} enables finer-grained comparisons, capturing partial improvements that \emph{EM} overlooks~\cite{Zhou2023}.

Their results show Tufano2022 achieves the highest \emph{ExactMatches} across all tasks among the baselines.
Among pre-trained transformers, \emph{CodeT5} outperforms all others on \emph{CodeRef1F+L} and \emph{CodeRef2F}, while Tufano2022 remains strongest on \emph{ComGenF}.
Finally, they demonstrate that rankings can shift when using \emph{EditProgress}.
For example, in \emph{CodeRef2F}, AutoTransform ranks highest in \emph{EP} despite not being among the top three in \emph{EM}.

\paragraph*{SMILER: Cao et al. (2024)~\cite{Cao2024}}
addresses the \emph{CodeRef1F} and \emph{CodeRef2F} tasks at method level granularity.
It extends an encoder decoder transformer with a conditional variational autoencoder (CVAE) to inject Gaussian noise into the generation process and increase code diversity~\cite{DBLP:journals/corr/abs-2305-00980}.
Despite its recent publication, the study uses the older and smaller Tufano 2021 dataset~\cite{Tufan2021} with \textapproxnew~17k instances and uses its model as a baseline.
A second baseline is a custom sequence to sequence model built with \emph{opennmt-py}.
Metrics include \emph{EM}, \emph{BLEU-4}, and \emph{LEV}.
At TOP-1 for \emph{CodeRef2F}, \emph{SMILER} achieves 5.93\% \emph{EM}, outperforming Tufano (2.91\%) and Seq2Seq (3.2\%).
For \emph{CodeRef1F}, it again leads with 17.68\% \emph{EM}, followed by Tufano (12.16\%) and Seq2Seq (13.26\%).
For \emph{BLEU-4} and \emph{LEV}, the paper reports mean, median, and variance, but no clear ranking emerges and no significance testing was conducted.

\subsection{Retrieval Models\label{sec-methods-retrieval}}
\paragraph*{DeepCodeReviewer: Gupta et al. 2018~\cite{Gupta2018}}
addresses the \emph{ComGenF+L} task by learning associations between new code snippets and corresponding review comments based on historical data.
The model retrieves relevant comments from a knowledge base using separate embeddings for upper, current, and lower lines of code.
LSTMs generate these embeddings, followed by fully connected layers computing code-comment relevance scores.
Negative examples were constructed by randomly pairing code with unrelated comments.
After filtering, the dataset contained \textapproxnew~31k positive instances, with experiments exploring various positive-to-negative ratios.
Best results were achieved with a 1:5 ratio.
Evaluation compared \emph{DCR} to a tf-idf logistic regression baseline.
Based on the precision-recall curve and relevance thresholds, \emph{DCR} achieved 26\% \emph{AUC}\footnote{The paper states an \emph{AUC} of 26. We assume this must be a percentage.}, a 62.5\% improvement over the baseline.
However, reaching 10\% recall required accepting over 60\% false positives.

\paragraph*{RSharer/RSharer+: Guo et al. 2019/2020~\cite{Guo2019,Guo2020}}
address the \emph{ComGenF} task at lines-level fragment granularity, aiming to share review comments across projects using code clone detection techniques.
\emph{RSharer} and \emph{RSharer+} combine AST-based deep learning with a ground knowledge base of \textapproxnew 78k code-comment pairs collected from StackOverflow, GitHub, Bitbucket, and SourceForge.
AST nodes are embedded using Word2Vec and aggregated into full AST embeddings.
A CNN-based clone detector compares embeddings and assigns similarity scores to rank candidate fragments and their associated comments.
The clone detector is trained on \emph{BigCloneBench}~\cite{Svajlenko2014}, which defines clone types from syntactic (Type-1) to semantic (Type-4), each mapped to a similarity score.
In clone classification, \emph{RSharer} outperformed prior models~\cite{Saini2018,Li2017a}.
For comment sharing, researchers generated the top-3 suggestions for 200 GitHub fragments.
Human annotators found at least one reasonable comment in 97 cases, compared to 68 for the \emph{CCLearner} baseline.
\emph{RSharer+} extends this with a semi-supervised denoising autoencoder, increasing this to 113 cases~\cite{Guo2020}.
It also classifies comment types and extracts topics using an LDA model~\cite{10.5555/944919.944937}.

\paragraph*{CORE: Siow et al. 2020~\cite{Siow2020}}
addresses the \emph{ComGenC} task at line-level fragment granularity, predicting relevant review comments for code changes.
\emph{CORE} learns from change-review pairs using two-level embeddings at the word and character levels, combined through a multi-attention mechanism to highlight key features.
Attention vectors are used to compute relevance scores between a code change and candidate reviews.
CORE used a custom GitHub dataset of 57,260 change-review pairs, with a 7:0.5:2.5 training, validation, testing split.
During training, each change was paired with five random comments to provide negative samples.
In terms of \emph{REC} and \emph{MRR}, \emph{CORE} outperformed both a TF-IDF baseline and DeepCodeReviewer~\cite{Gupta2018}.
The gains are attributed to the hybrid embedding design and multi-attention mechanism, which improve semantic matching and mitigate out-of-vocabulary issues.

\paragraph*{CommentFinder: Hong et al. 2022~\cite{hong2022commentfinder}}
addresses the \emph{ComGenF} task at method-level fragment granularity, framing comment generation as an information retrieval problem.
Methods are represented as both token sequences and bag-of-words token vectors, enabling construction of a ground knowledge base of \textapproxnew~134k instances with associated review comments.
To retrieve candidate comments, a new method’s token vector is compared using cosine distance to identify the ten closest instances.
These are re-ranked using \emph{Gestalt Pattern Matching}, which considers token order and yields more meaningful similarity scores than cosine similarity.
Evaluation used the Tufano et al. 2022 dataset, with the baseline model’s training split serving as the retrieval database and the test split for querying.
\emph{CommentFinder} outperformed the transformer baseline with TOP-1 \emph{EM} gains of 32\%, TOP-10 gains of 51\%, and \emph{BLEU-4} improvements of 3\% (TOP-1) and 28\% (TOP-10).
The results demonstrate the potential effectiveness of retrieval-based methods and support \emph{CommentFinder} as a competitive baseline for future work.

\paragraph*{RevCom: Shuvo et al. (2023)~\cite{Shuvo2023}}
addresses the \emph{ComGenC} task at line-level granularity using diff-level fragments.
The method extends retrieval-based comment generation by applying BM25~\cite{10.5555/188490.188561} and incorporating structured metadata such as file paths and import statements into unified instance representations.
Instances are tokenized using byte pair encoding (BPE), and vectorized with TF-IDF and custom Word2Vec embeddings, though the instance-level aggregation method for Word2Vec is not specified.
To generate comments, the 10 most similar instances are retrieved via BM25 and re-ranked using \emph{Gestalt Pattern Matching}.
Evaluation used a filtered dataset of 56k instances from eight large Java and Python GitHub projects, split 70\%/30\% into retrieval and test sets.
The authors ensured that comments from the same original change were not split between sets to avoid target leakage.
Metrics include \emph{BLEU-4}, \emph{EM}, and, to our knowledge, the first use of an embedding-based metric for semantic similarity built on \emph{SBERT} and \emph{stsb-roberta-large}.
\emph{RevCom} achieved a TOP-1 \emph{BLEU-4} of 14.84\% intra-project and 8.30\% inter-project.
It outperformed \emph{CommentFinder} by 49.5\% and \emph{CodeReviewer} by 23.57\% in TOP-10 \emph{BLEU-4}, with statistically significant gains across all metrics.
An ablation study showed that diffs had the greatest impact, followed by library imports and file paths.
Word2Vec yielded only minor improvements over TF-IDF.

%% file: content/results.tex
\section{Discussion\label{sec-discussion}}
Having addressed RQ1: MCR Tasks through our survey in Section \ref{sec-survey}, we now turn to RQ2: Metrics, RQ3: Results, and RQ4: Validity. 
In the following, we present and discuss our findings based on the data aggregated from the surveyed publications.

\subsection{RQ2: Metrics\label{sec-discussion-metrics}} 
First we explain our findings regrading the prevalence of metrics for different learning tasks, from which we derive recommendations for future research.
\begin{figure}[h]
	\includegraphics[width=\linewidth]{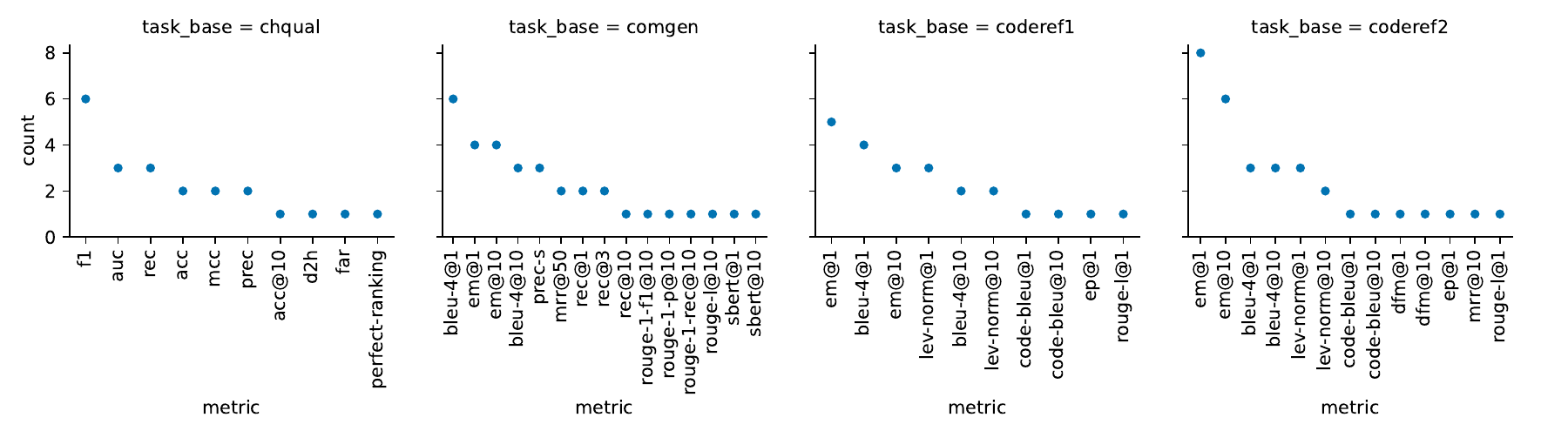}
	\caption{Frequency of metric usage for different learning tasks.\label{fig-metric-stats}}
\end{figure}

\subsubsection{ChQual}
Most of the surveyed papers treat \emph{ChQual} as a binary classification task, so we start with these.
Expectedly, the metrics that we saw most are the standard metrics for classifiers, i.e. Accuracy (\emph{ACC}), Precision (\emph{PREC}), Recall (\emph{REC}), F1-Score (\emph{F1}) and the Area-under-Curve metric for the Receiver-Operating-Characteristic (\emph{AU-ROC}).
We consider these metrics to be common knowledge in the computer science community, and we will not explain them further.
Of these metrics, \emph{F1} (harmonic mean of \emph{PREC} and \emph{REC}) was the most frequently used metric, probably due to the desire to have a single score that balances false positives and true positives.
The other standard metrics appeared half as frequently and their concrete frequency ranking is likely not expressive for the small number of samples.
For \emph{AUC}, we found that two articles specified that they used \emph{AUC-ROC} ~\cite{Shi2019, Wu2022}, while one article did not specify which flavor of AUC they used and there is also no code available ~\cite{Li2019}.
In addition to \emph{AUC-ROC}, there are other metrics such as AUC-PR, which is arguably preferable for imbalanced data.
Therefore, the distinction is very relevant for \emph{ChQual}, where in practice negative instances outweigh positives.
Beyond these standard classification metrics, we found a number of more specific metrics, Distance-to-Heaven (\emph{D2H}), False-Alarm-Rate (\emph{FAR}), and Matthews-Correlation-Coefficient (\emph{MCC}).
\emph{FAR} and \emph{D2H} are conceptually related to precision and F1-Score, and beyond giving the formula, the respective article does explain why these metrics in particular were chosen ~\cite{Hong20221034}.
\emph{MCC}) is also based on the confusion matrix, as it takes into account the numbers of true/false positive/negative instances.
Compared to \emph{F1}, it equally weights all four sectors of the confusion matrix and is therefore more suited for tasks with equal class importance.
Neither of the two articles using \emph{MCC} give specific reasons about why they chose this metric or discuss its results in contrast with \emph{F1} ~\cite{Staron2020513,Wu2022}.
Accuracy-at-10 (\emph{ACC@10}) and Perfect-Ranking are metrics for ranking-based \emph{ChQual}.
(\emph{ACC@10}) measures the proportion of instances (here: files) where one of the top 10 ranked fragments actually exhibited a comment in the ground truth~\cite{Hong20221034}.
Perfect-ranking measures the proportion of instances (here: change sets) where all fragments with ground-truth comments were ranked above all of the fragments without comments~\cite{Hellendoorn2021}.

\finding{ChQual Recommendations}{
Simple classification approaches should focus on the simple and commonly understood \emph{ACC, PREC, REC, F1} metrics. 
If a more nuanced picture of models' performance is desirable, in particular when the use-case involves ranking changes by criticality instead of a binary classification, the \emph{AUC} metrics and \emph{Perfect-Ranking} should be preferred.
Due to the highly imbalanced nature of \emph{ChQual} data, consider using \emph{AUC-PR} instead of \emph{AUC-ROC}.
}

\subsubsection{ComGen}
For \emph{ComGen} the overall most prevalent metrics, \emph{BLEU-4} and \emph{ExactMatches} were the first metrics proposed for generative MCR models~\cite{Tufan2021,Tufano2019}.
The former was originally called \emph{PerfectPredictions}, however, we believe the naming by Li et al. should be preferred, as it does not carry the ambiguity of the word 'perfect' ~\cite{Li20221035}.
The \emph{BLEU} family of metrics was originally intended for the automatic evaluation of machine translation algorithms, by comparing candidate translations to sets of reference translations~\cite{papineni2002bleu}.
Simply put, it computes the (precision) overlap of n-grams, typically 4-grams, appearing in the candidate translation vs. references translations and includes extra factors for discounting very short or repetitive sentences, since, compared to longer sentences, these are generally more probable to have a high n-grams overlap with the references.
\emph{BLEU} scores range from 0 (worst) to 1 (best) and are often expressed as percentages. 
Beyond the n-gram order, there are a number of important details, e.g. the choice of lexer, corpus-based vs. sentence-based computation, smoothing methods, specific implementation, etc.
To ensure a valid comparison between multiple approaches, researchers should report details on these aspects to be sure that others can replicate the same configuration of the metric. 
In our survey, details on these aspects were scarce but could mostly be derived from the replication packages.
This challenge was previously reported for other domains.
Therefore, libraries such as \emph{sacrebleu} facilitate reporting \emph{BLEU} results in a reproducible way ~\cite{post-2018-call}.
Furthermore, there are no generally recognized guidelines on qualitative interpretation of BLEU scores for the MCR domain, although increasing scores were shown to align with increasing human evaluations for the machine translation domain ~\cite{reiter-2018-structured}.
While \emph{BLEU} attempts to quantify the textual similarity between candidates and references, \emph{ExactMatches} only considers the generated comments identical to the ground truth to be correct.
Thus, it is a much simpler metric which is easier to compute and less likely to be misinterpreted.
Beyond \emph{BLEU} and \emph{ExactMatches}, several additional metrics were used, although less frequently.
Two approaches used the \emph{ROUGE} metric, which is also an n-gram overlap metric and therefore similar to \emph{BLEU} but can include both precision overlap and recall overlap, depending on its configuration ~\cite{Li2022a,Yin2023}.
Yin et al. thus state that it focuses on 'completeness' instead of 'accuracy', potentially making it a reasonable complement to \emph{BLEU}~\cite{Li2022a,Yin2023}.
The variant \emph{ROUGE-L} makes use of the length of the longest common subsequence between the candidate and reference instead of the fixed-length n-gram statistics~\cite{DBLP:conf/acl/LinO04}.

A major drawback of this type of lexical similarity metrics is their inability to capture semantically equivalent comments. 
On the one hand, \emph{BLEU, ROUGE} and especially \emph{ExactMatches} will typically score a candidate comment badly if it expresses the same meaning as the ground truth, but using a different wording.
On the other hand, the metrics will score a candidate highly, even if it expresses the complete opposite of the ground truth, if this change in semantics is due to a small change in wording, e.g. a single insertion of the word 'not'.
In terms of semantic similarity metrics, we found only one application of \emph{Sentence-BERT Similarity (SBERT)}~\cite{Shuvo2023}.
This metric computes the cosine distance between sentence vector embeddings that are derived for candidate and ground-truth comments using the Sentence-BERT framework~\cite{reimers-gurevych-2019-sentence}.
While a promising approach, it is important to understand that sentence similarity is a domain-specific concept, e.g. the similarity of code review comments differs even from the similarity of stack-overflow questions. 
Thus, the embedding model would benefit from fine-tuning.
As there currently exists no appropriate dataset, there is no information on how good the metric used by Shuvo is, or how well any other metrics using sentence embedding models for other domains actually capture the relevant aspects of similarity between code review comments.

Three approaches were used, what we call the \emph{SubjectivePrecision} (PREC-S) metric which is based on human judgments of the comments generated~\cite{Guo2019, Guo2020, Gupta2018}. 
Although different names, e.g.'reasonable (shared) reviews', are used for the metric, the idea is the same: a human decides whether a generated comment is valid or not.
While this metric has the advantage of not requiring any ground-truth comments, it is inherently subjective, or multi-subjective in case multiple humans are asked, and thus hard to reproduce.

Since most models can be used to generate a top-k sample of candidates instead of just one comment, all of the above metrics can be used and were used in different \emph{METRIC@K} configurations.
In this case, \emph{ExactMatches@k, BLEU@k, ROUGE@k}, and \emph{PREC-S@k} are interpreted as the average of the best scores that the model achieved within each top-k sample of candidates.
Slightly different @k-semantics apply to the next two metrics, \emph{Recall (REC@K)} and \emph{MeanReciprocalRank (MRR@k)}, since they are used to evaluate ranking models, which suggest historical comments from a database by ranking them according to some similarity function.
The \emph{MeanReciprocalRank (MRR@k)} gives the average inverse index (rank) of the first correct comment in the sets of retrieved and ranked k-candidates.
So in case a model suggests multiple valid comments in a sample of k candidates, \emph{MRR@K} ignores all except for the first one.
In MCR automation, Gupta et al. were the first to use \emph{REC@K} to measure ranking performance~\cite{Gupta2018}.
According to their definition, which was later referenced by Siow et al. ~\cite{Siow2020}, \emph{REC@K} measures the proportion of instances for which at least one valid comment was suggested in the sample of the top-k candidates.
The \emph{REC@K and MRR@K} are therefore complementary metrics.
Examining them at different levels of \emph{k} could give insight into how many candidates a developer would need to examine in practice to get satisfactory suggestions~\cite{Siow2020}.
In a more general setting, where multiple valid comments can be imagined for a code change, the standard definition of recall in information retrieval would be applicable. 
In this case, the metric measures the proportion of relevant comments retrieved for a change versus all relevant comments for that change, averaged across all changes in the dataset~\cite{Rijsbergen}.

\finding{ComGen Recommendations}{Despite of their known shortcomings \emph{BLEU-4} and \emph{EM} have been the default \emph{ComGen} metrics and researchers should report them until a new metric is shown to be superior.
	We suggest reporting them at least @1 and @10.
	The primary challenge for \emph{ComGen} metrics is that so far, no metric can adequately capture \emph{semantic} similarity, and so far little research has been published on this topic.
	Embedding-based metrics like \emph{SBERT}, in particular using tailored embedding models that do not yet exist, are a promising avenue for further research and should be used complementary to \emph{BLEU-4} and \emph{EM}.
	Until then, human-based evaluations are important to add a qualitative dimension to understanding models' performance.
	To this end, we suggest further standardization of such experiments in order to achieve higher comparability.}

\subsubsection{CodeRef1/CodeRef2}
The most common metrics for both \emph{CodeRef} tasks are again \emph{ExactMatches}, \emph{CodeBLEU} and \emph{BLEU-4}.
Since the \emph{BLEU} metrics were designed to evaluate the quality of natural language translations, Tufano et al. proposed to use the \emph{CodeBLEU} metric instead for the \emph{CodeRef} tasks, which leverages Abstract Syntax Trees and dataflow to get a source code specific quantification of similarity~\cite{Tufano20222291, DBLP:journals/corr/abs-2009-10297}.

Regarding \emph{CodeBLEU}, Tufano reports on an important issue:
Program changes are on average very small, influencing only a small number of lines or tokens with regard to the original source code. 
Thus, there is a large overlap between the original code and the changed code in the ground truth.
In consequence, any \emph{CodeRef} model that makes a different, but small change to the original code is still likely to achieve high average \emph{CodeBLEU} scores, in Tufano's case averaging above 80\% median \emph{CodeBleu}. 
However, since very small differences in source code lead to very different behavior in practice, Tufano et al. suggest focusing on \emph{ExactMatches} as a more precise way of quantifying correct code refinements performed by a model~\cite{Tufano20222291}.
We argue that the same limitations apply to standard \emph{BLEU-4}, which was used to measure \emph{CodeRef} performance by other groups and also by Tufano in earlier work ~\cite{Tufan2021,Cao2024,Li20221035,Lu2023}.

Similarly to \emph{CodeBLEU}, Lin et al. used a custom metric called \emph{DataFlowMatches (DFM)}, but the two metrics are not explicitly compared to explain the benefit of one over another.

Beyond these metrics, some groups used a normalized distance \emph{Levenshtein (LEV)} instead of or complementary to \emph{BLEU} metrics.
\emph{Levenshtein} distance is an edit distance that counts the number of insertions, deletions, or replacements of characters that are necessary to transform one string into another.
Therefore, a small metric indicates a high similarity.
Since the range of values is proportional to the length of the compared strings, the surveyed papers reported scores that were normalized by the maximum length, resulting in scores between 0 for identical strings and 1 for completely different strings.
For us, the \emph{Levenshtein} distance has the same interpretability issues as \emph{BLEU} metrics.

The \emph{EditProgress (EP)} metric, recently adopted by Zhou et al. for code refinement and introduced in Section \ref{sec-survey}, has been used only once so far~\cite{Zhou2023}.
It is conceptually related to \emph{Levenshtein}, but essentially extends it with the comparison to an \emph{implicit, naive no-change baseline model}.
Therefore, we believe that it may be advantageous with regard to interpretability issues described in the previous paragraphs, as it gives basic insight into whether at least partial progress toward the ground truth is achieved.
We also agree with the authors' notion that it should allow for a more fine-grained comparison of models, especially compared to \emph{ExactMatches}.

Finally, only one paper used \emph{Rouge-L} (see \emph{ComGen}) and the \emph{MeanReciprocalRank (MRR)} of the first exact match among the top-10 predictions to evaluate \emph{CodeRef} performance~\cite{Yin2023}.

\finding{CodeRef Recommendations}{Following the reasoning of Tufano, \emph{EM} should remain the primary metric for \emph{CodeRef} tasks.
Between the other metrics, we suggest using standard \emph{BLEU-4}, due to the availability of transparent standard implementations like \emph{sacrebleu}, which should allow better reproducibility.
We suggest to report metrics at least @1 and @10.
Although source code-specific similarity metrics like \emph{CodeBLEU} and \emph{DFM} appear reasonable (to us), they should be discussed in contrast to the primary metrics to better understand their contribution.
Finally, especially if there is no accepted guideline for qualitatively interpreting a particular metric, including a measurement of partial progress, as demonstrated by \emph{EditProgress}, should be considered to ensure that the models actually make progress toward the ground truth.}

In conclusion, throughout our survey, we experienced several recurring tasks-independent themes with respect to the use of metrics.
In particular, we found that of a total of 48 task-metric combinations, 22 were unique to their original paper.
Also, while we did not track statistics for this, we had to search for, and often did not find, details that would be necessary to accurately reproduce many metrics.
Therefore, we derive the following recommendations.

\finding{General Metric Recommendations}{
\begin{itemize}[left=0pt]
	\item Focus on established standard metrics
	\item If you introduce a metric, use it in a complementary way, explain its contribution and discuss it in contrast with other metrics 
	\item Be precise about metric flavors, normalization, averaging, implementation and all other aspects required to understand and reproduce the metric
	\item Prefer commonly-available, well-used metric implementations
	\item Supply metric computation as part of your replication package
\end{itemize}}

\subsection{RQ3: Results\label{sec-discussion-results}} 
In the following subsections, we put the results reported by the different articles in our survey into a unified, bigger picture to convey an understanding of the state-of-science in MCR automation.
We treat the learning tasks individually and use a grid of diagrams, one for each metric in Section \ref{sec-discussion-metrics}.
Note that we exclude metrics that were only used once, since with only a single data point no comparison is possible.
We further split the visualization by dataset into four categories, \emph{Paper Specific}, \emph{Tufano 2021}, \emph{Tufano 2022} and \emph{CodeReviewer} ~\cite{Tufan2021,Tufano20222291,Li20221035}.
The latter three datasets belong to the corresponding papers and are, so far, the only datasets we found to have been reused by different researchers.
Therefore, the metric values within each of these categories have the highest comparability.
The \emph{Paper Specific} category contains results that were produced on paper-specific datasets that were not shared with any other approaches.
Thus, the comparability within this category is more limited; however, the general range and distribution of values still give an indication of what kind of performance has been achieved so far.

\subsubsection{ChQual}
\begin{figure}
	\includegraphics[width=\linewidth]{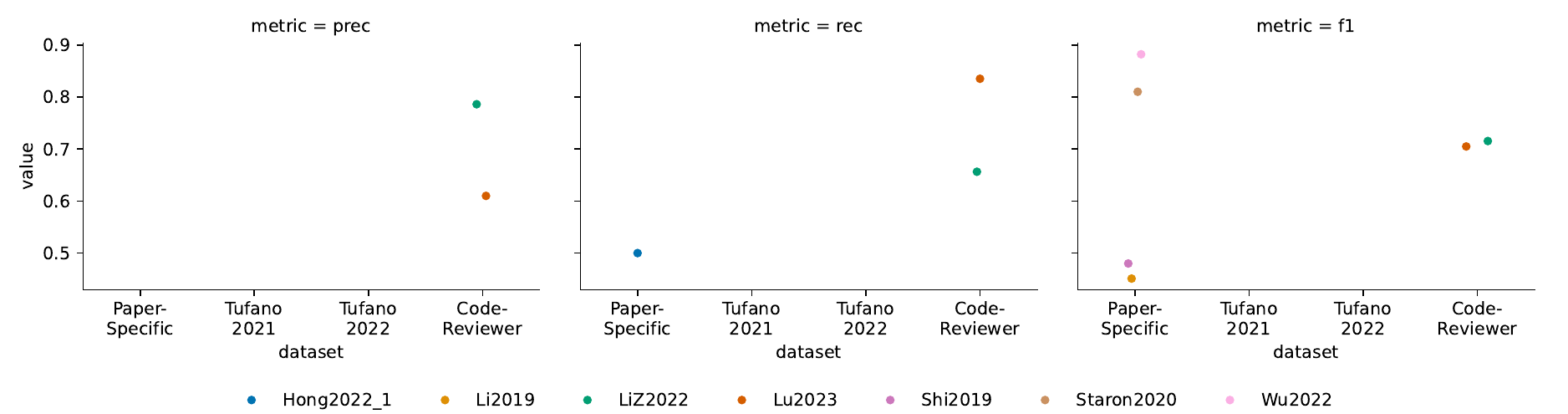}
	\caption{Reported results of ChQual approaches. Paper-Specific category shows results for different datasets that were only used by exactly one paper. Note that the\emph{CodeReviewer} dataset was the only dataset that was re-balanced both on training and validation data.\label{fig-results-chqual}}
\end{figure}

With regard to Section \ref{sec-discussion-metrics}, for \emph{ChQual} we focus on \emph{Precision}, \emph{Recall} and \emph{F1}, and also omit \emph{Accuracy} since it is not suitable for imbalanced data.
The diagrams for all metrics can created with our replication package.

Figure \ref{fig-results-chqual} gives an overview of the results reported for \emph{ChQual}.
We found that the only dataset that was used by more than one group is the \emph{CodeReviewer} dataset~\cite{Li20221035}.
Here we have two sets of metric values, first by Z. Li et al. themselves, and more recently by Lu et al. from their 2023 \emph{LLAMA-Reviewer} Paper.
Taking into account \emph{F1}, \emph{CodeReviewer} scored 0.705 and \emph{LLAMA-Reviewer} 0.715, a difference that is likely not relevant in practice.
Compared with the other metrics, we see primarily that \emph{CodeReviewer} is more \emph{recall}-oriented, while \emph{LLAMA-Reviewer} focuses on \emph{precision}.
In contrast to all other data sets, it is important to note that \emph{CodeReviewer's} dataset is artificially balanced, both in training and in validation data.
We consider this to be an important fact when comparing these results with those of the other, paper-specific datasets. 

For the paper-specific datasets, we essentially have two groups, Wu et al. and Staron et al., as well as H. Li et al. and Shi et al.
The former reported the highest F1 scores above 80\%~\cite{Wu2022, Staron2020513}, while the latter reported scores below 50\% ~\cite{Shi2019, Li2019}.
Considering the highly imbalanced nature of the data and especially comparing them to the results achieved on balanced \emph{ChQual} data, these results appear to be surprisingly high.
For Staron et al. this may be explained by the very small dataset of only 407 instances, featuring only 73 positive instances.
In both cases further research is advisable to verify these results

\finding{ChQual Findings}{
Due to the small number of samples and the huge spread between the best and worst reported results in \emph{ChQual}, we cannot see a general trend for learning task. 
However, since the best results were only achieved on paper-specific datasets, one of which was extremely small, we suggest that these approaches should be evaluated on different datasets to confirm their external validity.}

\subsubsection{ComGen\label{sec-discussion-metrics-comgen}}
\begin{figure}
	\includegraphics[width=\linewidth]{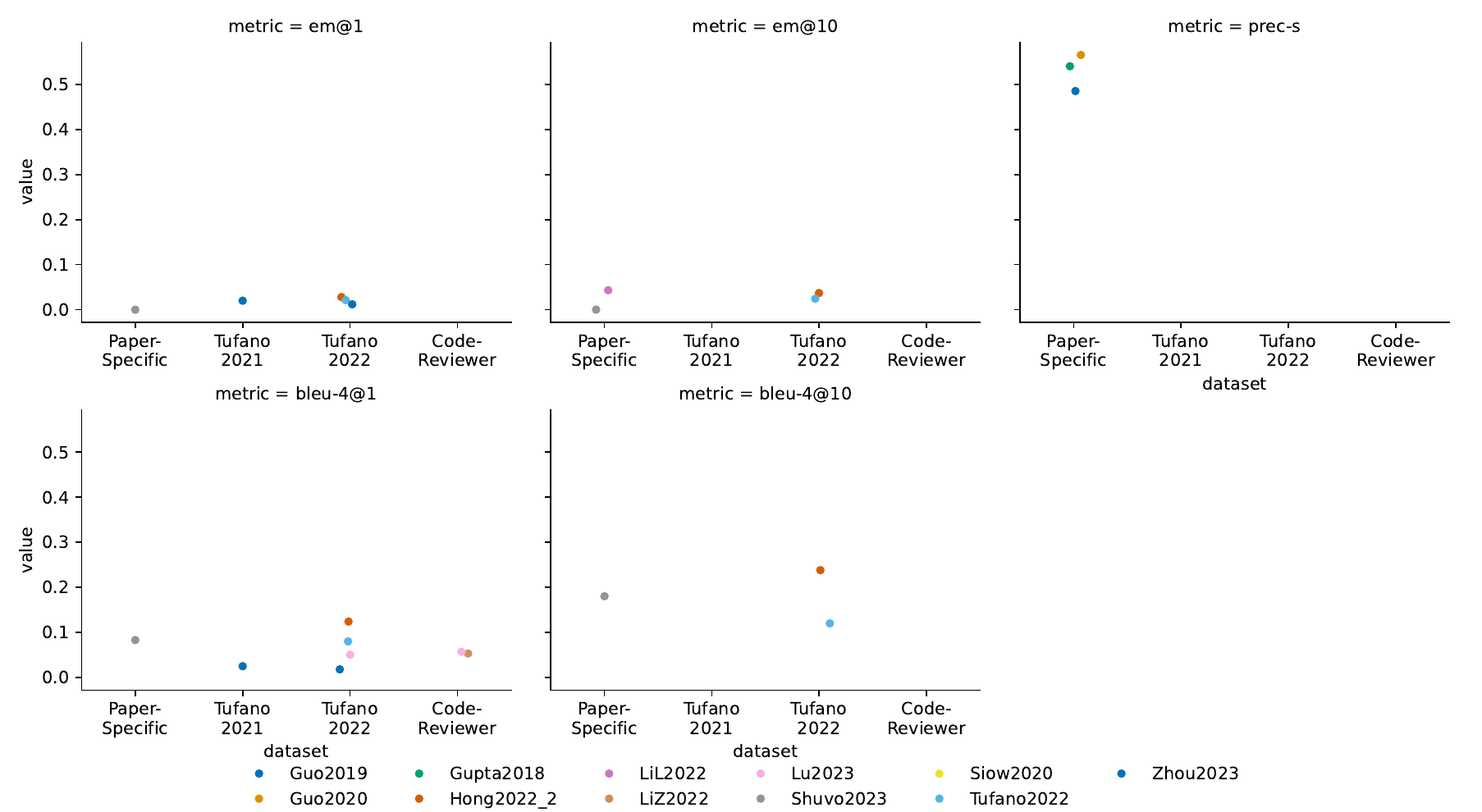}
	\caption{Reported results of ComGen approaches. Paper-Specific category shows results for different datasets that were only used by exactly one paper.\label{fig-results-comgen}}
\end{figure}

As explained in Section, \ref{sec-discussion-metrics} for \emph{ComGen} we focus on the \emph{ExactMatches}, \emph{BLEU-4} and \emph{SubjectivePrecision} metrics.
The diagrams for all metrics can be found in our replication package.
Figure \ref{fig-results-comgen} gives an overview of the results reported for \emph{ComGen}.

For both \emph{ExactMatches} and \emph{BLEU-4}, the reported results are consistently very low across all the approaches surveyed.
\emph{EM@1} scores range from 0.003\% to 3\% and the \emph{EM@10} scores range from 0.01\% to 4\%.
This underscores the assumption that no \emph{ComGen} model is likely to do well at exactly predicting human code review comments, even when considering Top-10 results.
The best results for both \emph{EM@1} and \emph{EM@10} are those reported by Hong et al. ~\cite{hong2022commentfinder}.
Since their \emph{CommentFinder} retrieves comments from a database of historical changes and comments, it is plausible that \emph{CommentFinder} could achieve higher \emph{EM} scores than generative models for very common comments.

Looking at \emph{BLEU-4@1}, the scores range from 1\% to 12\% and for \emph{BLEU-4@10} from 2\% to 23\%.
Although there are no guidelines for qualitatively interpreting \emph{BLEU} scores in the MCR domain, for general machine translation, Google's translation API documentation states that scores below 10\% are "almost useless", for 10\% to 19\% the meaning is "hard to grasp" and that upward of 20\% bleu, translations become "acceptable but still have significant grammatical errors" ~\cite{googlebleu}.
If these guidelines are in any way applicable to the similarity of code review comments, which are mostly technical English, state of science \emph{ComGen} approaches could thus be just on the verge of becoming useful in terms of \emph{BLEU}.

The results reported by the articles using the \emph{SubjectivePrecision} metric based on human evaluation of the comments range from 48\% to 56\%.
Compared to the before metrics, this indicates that although current approaches struggle to generate comments syntactically similar to or identical to a ground truth, they may be capable of generating reasonable comments in many cases.

\finding{ComGen Findings}{
The results in terms of \emph{EM} and \emph{BLEU} indicate that the state-of-science models still need to improve significantly to become applicable in practice.
Since neither of these metrics adequately captures semantic similarity, complementing the results of automated metrics with human evaluations such as \emph{SubjectivePrecision} is important to get a better understanding of performance. 
Such studies found that approximately half of the comments generated appeared valid to the human inspector.}

\subsubsection{CodeRef1/2}
\begin{figure}
	\includegraphics[width=\linewidth]{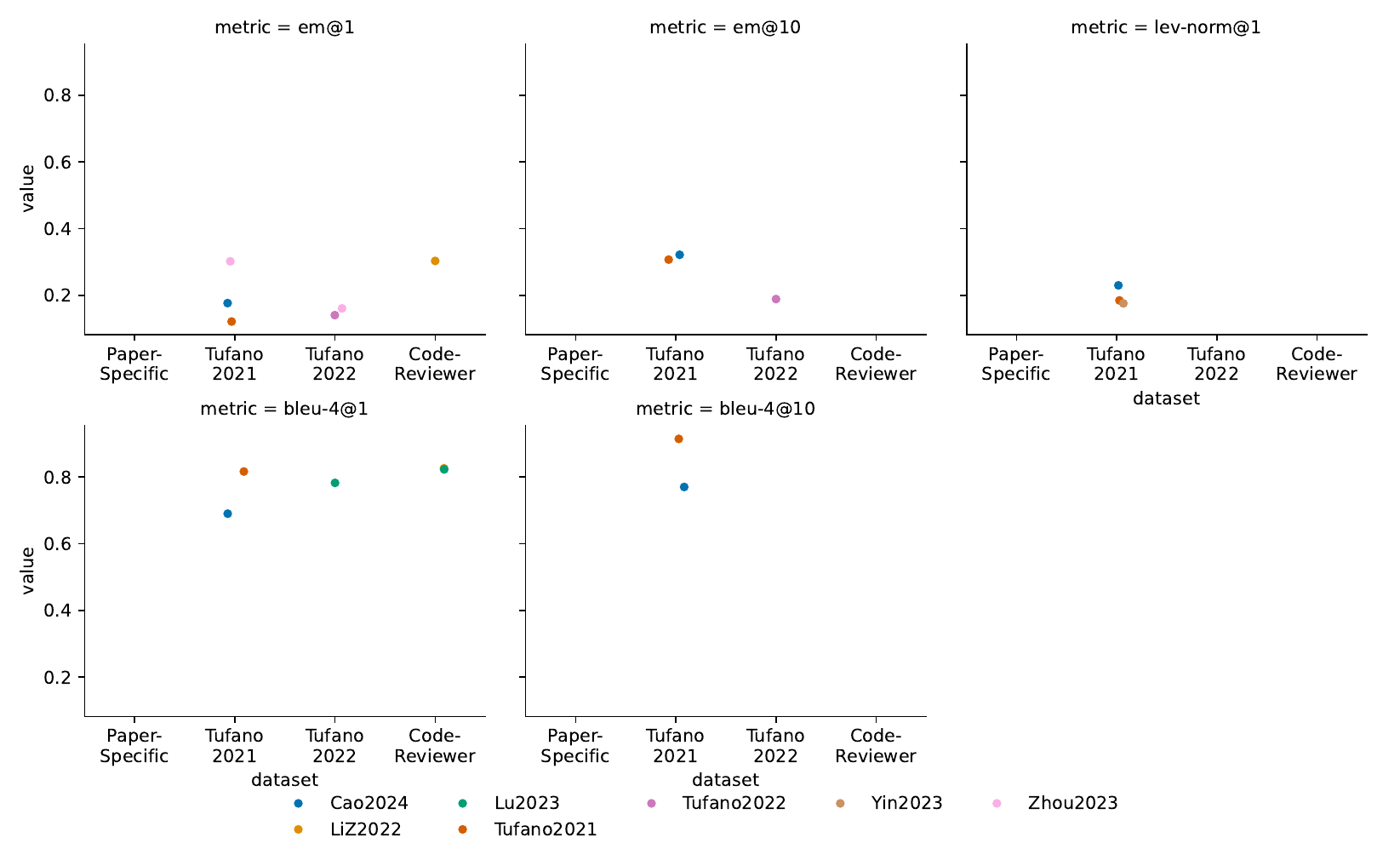}
	\caption{Reported results of CodeRef1 approaches. Paper-Specific category shows results for different datasets that were only used by exactly one paper.\label{fig-results-coderef1}}
\end{figure}

\begin{figure}
	\includegraphics[width=\linewidth]{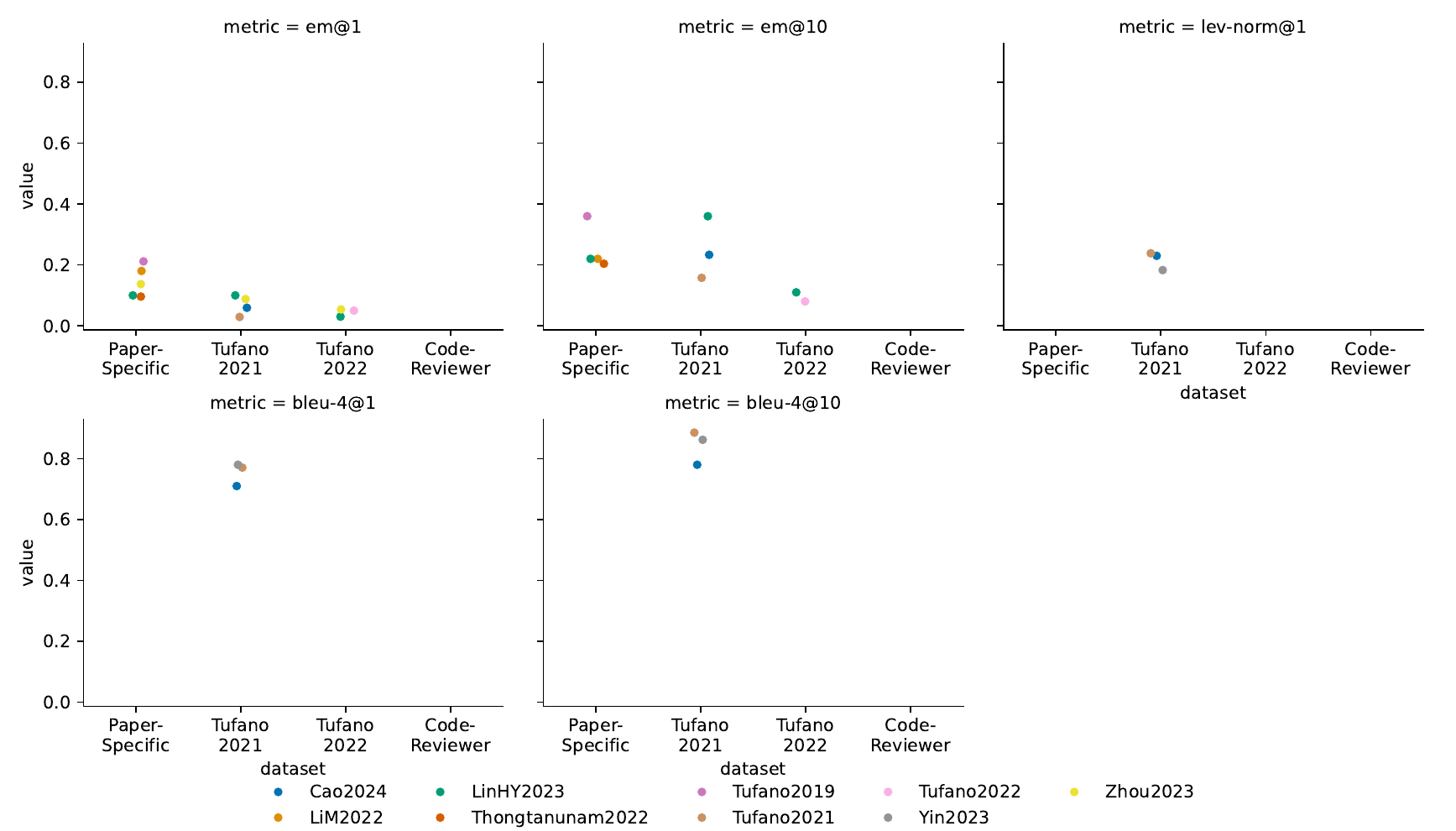}
	\caption{Reported results of CodeRef2 approaches. Paper-Specific category shows results for different datasets that were only used by exactly one paper.\label{fig-results-coderef2}}
\end{figure}

\begin{figure}
	\includegraphics[width=\linewidth]{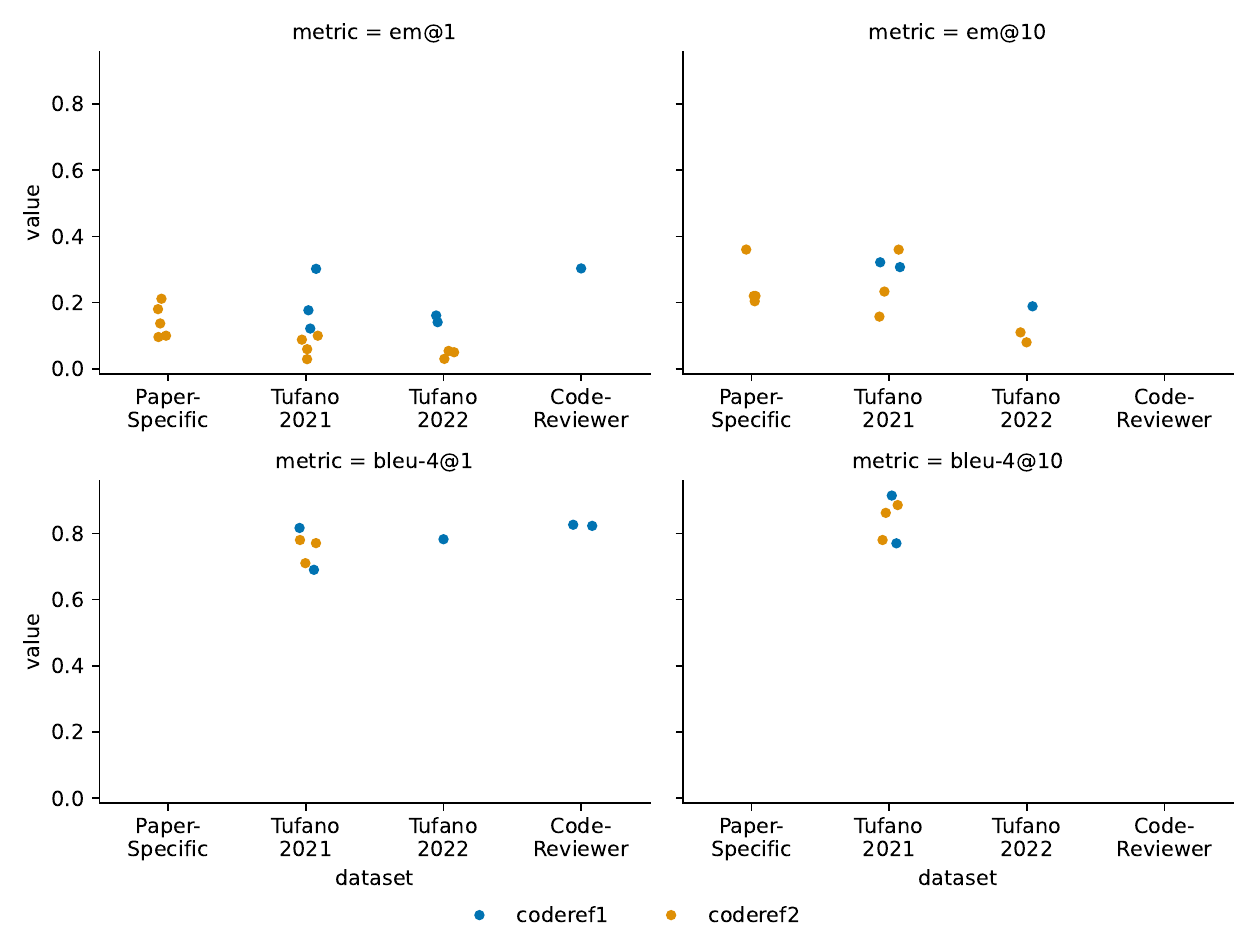}
	\caption{Comparison of reported results for CodeRef1 vs. CodeRef2 approaches. Paper-Specific category shows results for different datasets that were only used by exactly one paper. \label{fig-results-coderef1vs2}}
\end{figure}

With regard to Section \ref{sec-discussion-metrics}, for \emph{CodeRef} we focus on the metrics \emph{ExactMatches} and \emph{BLEU-4} and also present the results for \emph{Levenshtein}.
Since the source code-specific metrics, \emph{CodeBLEU} and \emph{DFM}, were only used by a single paper, they are omitted but can be generated using our replication package.

Figures \ref{fig-results-coderef1} and \ref{fig-results-coderef2} show the results individually for \emph{CodeRef1}, which includes comments in the refinement of the code, and for \emph{CodeRef2}, which learns to refine the source code without the use of comments.
Like with Figures \ref{fig-results-chqual} and \ref{fig-results-comgen} the color dimension identifies the results from the same paper across multiple datasets.

For \emph{CodeRef1}, across all datasets the reported results for \emph{EM@1} range from 12\% to 30\% and for \emph{BLEU-4@1} from 69\% to 82\%.
In both cases the best results were reported for the \emph{CodeReviewer} dataset.

For \emph{EM}, predicting the exact code change that resulted from a human code review in even 12\% of the cases appears considerable.
Assuming the results would generalize to other datasets, 30\% \emph{EM} as reported by Z.Li et al. could even provide an acceptable user experience when integrated into a code review platform.
What appears puzzling is the relatively small increase in performance when comparing the \emph{EM@10} scores, which range from 19\% to 32\%, especially the mere increase of 2\% for the model with the highest rating, \emph{CodeReviewer}~\cite{Li20221035}.
As one explanation could be over-fitting and/or target leakage, which could result in lacking diversity in the top-10 predictions, we believe this underscores the need for further validation on other datasets.

Source code is very different from natural language, in particular, much smaller changes to source code tend to have much higher impact on semantics.
Thus, regarding \emph{BLEU}, the guidelines mentioned in Section \ref{sec-discussion-metrics-comgen} to interpret the metric in the case of natural language are likely even less applicable.
Consequently, the \emph{BLEU} metric should only be used to relatively compare between models or with a suitable baseline.
In the simplest case, this could be a naive, no-change baseline that always predicts the original code without performing any modifications.
This would reveal whether the \emph{CodeRef} model is actually taking steps to improve the code.
Since none of the articles surveyed included such a baseline, for \emph{BLEU} we can thus only conclude that the reported performance of the model is fairly consistent in the range of 70\% to 80\%.
Looking at the expectedly higher \emph{BLEU-4@10} scores (77\% to 91\%) yields no further insights.

For the Levenshtein metrics, \emph{LEV-NORM@1} and \emph{LEV-NORM@10} the picture is similar to \emph{BLEU-4}.
Although the ranking between approaches differs, the general spread of results and interpretability issues are the same.

Regarding \emph{CodeRef2} there are significantly more approaches and thus more data-points, as well as some additional paper-specific datasets. 
Here, \emph{EM@1} ranges from 3\% to 21\% and \emph{EM@10} from 11\% to 36\%. 
Again, one of the two top-ranking models only exhibited a very small improvement from \emph{EM@1} to \emph{EM@10}, which should be further analyzed~\cite{Li2022}.
While this is not the case for the approach that reported the overall best results, its comparatively much higher performance could be due to the use of a very small and highly preprocessed dataset ~\cite{Tufano2019}.
Thus, we argue that the realistic performance for state-of-science approaches is around 10\% \emph{EM@1} and 20\%-30\% \emph{EM@10}.

Similarly to \emph{CodeRef1}, we believe that the \emph{BLEU-4@1} scores (69\% to 82\%) and the \emph{BLEU-4@10} scores (77\% to 91\%) do not provide any further insights.
The same is true for looking at Levenshtein.
Again, while the ranking differs slightly for \emph{LEV-NORM@1}, the only other metric used by more than one paper, this could well be a coincidence.

Finally, we are also interested in seeing whether there is a clear difference in the performance between \emph{CodeRef1} and \emph{CodeRef2} models, as can be intuitively expected due to the additional information available to the model.
Figure \ref{fig-results-coderef1vs2} shows a comparison of all reported \emph{CodeRef} results, where the color indicates the type of task, i.e. \emph{CodeRef1} or \emph{CodeRef2}.
As the figure shows, this does appear to be the case, especially for \emph{EM@1} and \emph{BLEU-4@1} for the \emph{Tufano2021/2022} datasets, however, there is too little data for the difference to significant.

\finding{CodeRef Findings}{
The results in terms of \emph{EM} are much closer to a practical applicability than for \emph{ComGen} or \emph{ChQual}.
However, there is still much potential for improvement, and existing as well as new approaches should be validated on a larger number of datasets to assert their true performance.
Based on the articles surveyed, we could not conclusively confirm the hypothesis that the additional comment information in \emph{CodeRef1} compared to \emph{CodeRef2}  leads to superior performance. 
Thus, further research is required on this topic.
Finally, we found that the qualitative interpretation of the \emph{BLEU} metric would require adequate guidelines or at least some simple baselines as a frame of reference.}

\subsubsection{Impact of Context and Focus}
We further investigated whether the context (change vs. fragment vs. review level) or focus (none vs. line vs. token level focus) showed any obvious impact on the results, as this was suggested, for example, by Z. Li et al ~\cite{Li20221035}.
Based on our collected data, we could not decisively observe this trend across multiple approaches in any of the MCR tasks.

\subsection{RQ 3: Validity\label{sec-discussion-validity}}
In this research question we analyzed the surveyed papers for explicit mentions of possible validity issues and challenges.
Clustering the results by internal, construct, and external validity, this section presents our findings and derived recommendations, so that other researchers can benefit from these insights on what to consider when engaging in MCR automation research.

\subsubsection{Internal Validity}
\paragraph{Correctness of implementations}
If the implementation of a model or experimental setup is incorrect, the observed effects may not be due to the intended method, but to implementation errors.
A special case is the re-implementation of baseline models if they were either unavailable or researchers were unable to run the original models.
Researchers should thus take care to thoroughly review and test their implementations and publish them to ensure they are available to their peers for validation and reuse.

\paragraph{Target Leakage \& Temporal Bias}
First, some articles reported mitigating the threat of target leakage by being careful not to split up related instances, such as multiple comments belonging to the same change, between training and validation sets.
Second, \emph{Temporal-bias} ~\cite{Jimenez2019} refers to an implicit leakage between training and evaluation sets. 
This leakage occurs because of the strong relationship between code fragments that are part of a shared evolution. 
It is an inherent challenge in learning MCR automation tasks ~\cite{Hong20221034}.
We found that only one of the surveyed publications explicitly addressed the topic ~\cite{Hong20221034}.
While many of the surveyed publications potentially address this challenge implicitly, for example by using project-level splitting between training and validation sets (~\cite{Li20221035}), some approaches do not provide detailed information on their splitting scheme for cross-validations or held-out-validations (e.g. ~\cite{Li2019, Shi2019}).
According to Hong et al. default random splitting is not an option in most cases due to the risk of temporal bias.
Hong et al. instead split datasets based on actual time sequences~\cite{Hong2022}.
For example, a version control system's commit history can be exploited to ensure that validation sets only include changes from commits that are strictly younger than those in the training set.

\paragraph{Training issues}
No or insufficient hyperparameter tuning may lead to validity issues, because one cannot know whether the observed performance of a model reflects the approach's true capabilities.
This is especially important with regard to baseline approaches that ideally are tuned to the same extent as the primary models.

Further, researchers reported that stochastic effects, for example different initialization of models' weights, can lead to differing results.
Thus, when feasible, models should be trained multiple times using different starting configurations.

\paragraph{Significance}
When comparing the performance of different models, especially for \emph{ComGen} and \emph{CodeRef} tasks, the results are distributions of metric scores like \emph{BLEU}, \emph{ROUGE} or \emph{F1}.
Comparing only averages of these distributions potentially loses a lot of information and, especially for small datasets and small performance differences, these can likely be due to the random sample of analyzed data.
To mitigate this risk, only 4 of the 18 surveyed \emph{ComGen} and \emph{CodeRef} publications used statistical significance testing ~\cite{Shuvo2023,Siow2020,Tufano20222291,Yin2023}, three of which also included a measure of effect size~\cite{Shuvo2023,Siow2020,Tufano20222291}.

\finding{Internal Validity Findings}{Researchers have addressed various challenges to internal validity and corresponding coping strategies, which we summarized above.
We believe that two points, in particular, require greater awareness: the concept of temporal bias and the use of statistical significance and effect size when comparing the results of different models and baselines.}

\subsubsection{Construct Validity}
\paragraph{Unsuitable Metrics}
As discussed in Section \ref{sec-discussion-metrics}, metrics such as \emph{BLEU} are very limited in their ability to capture semantic similarity.
Thus \emph{BLEU} as a construct does not accurately capture the intended concept of semantic similarity with regard to a ground truth.

Some approaches used proxy constructs, such as comment sentiment, for change quality~\cite{Staron2020513}.
When using such proxies, it is important to discuss possible implications.
For example, the dataset by Staron et al. only features changes that did, in fact, receive a comment before further discriminating these changes by sentiment. 
Consequently, there is a risk of an inherent selection bias resulting in the model performing differently in practice, when it is not known ahead of time whether a change is commented.

\paragraph{Appropriate Baselines}
The use of appropriate baselines in experimental evaluations is crucial to interpreting results, especially when comparing complex models such as transformers. 
Referring to Table \ref{fig-survey-overview} in our survey, we found that 15 out of 24 publications either used only transformers and other neural 'black-box' models as baselines or reported no baselines at all.
While differences between models and baselines may be statistically significant in some cases, we found that such comparisons often provide little context since the real-world performance of the baselines is also unknown. 
Without a known and understood reference point, it becomes difficult to gauge whether models' predictions align with real-world expectations or merely exploit patterns in the data to optimize metrics.

For \emph{ChQual} models, we suggest using simple classifiers based on synthetic features as baselines, for example by exploiting the notion that the complexity of changes is correlated with the likelihood of being commented.
We present an example of such a model in the following Section \ref{sec-experiment}, to underscore the importance of the frame of reference it provides to understanding the performance of more complex models.

For \emph{ComGen} models, we suggest using retrieval models such as \emph{CommentFinder} as a baseline~\cite{hong2022commentfinder}.

For \emph{CodeRef} models, we propose including at least a naive no-change baseline that simply retains the original code.
This baseline can be used to tell whether a model actually made an improvement to the code or just "moved sideways", i.e. made a modification that neither improved nor worsened the code with regard to the ground truth. 

\finding{Construct Validity Findings}{We identified several challenges to construct validity, two of which we consider particularly important.
First, the lack of semantic similarity metrics for code review comments is likely the biggest impediment to advancing comment generation.
Designing sentence embedding models tailored to code review offers a promising solution.
Second, the majority of analyzed articles used only 'black-box' models as baselines or provided no baselines at all, severely limiting the qualitative understanding of the attained results.
To address this, we proposed concrete baselines for each learning task.}

\subsubsection{External Validity}
\paragraph{Datasets and Data Collection}
Researchers reported a number of threats to validity with regard to datasets: evaluation on small datasets can lead to results that do not generalize to larger, real-world scenarios. 
Similarly, studies that use only a small number of datasets, or a single dataset, risk drawing conclusions based on dataset-specific characteristics rather than general trends. 
Limited language diversity (e.g., focusing only on Java) restricts the applicability of findings to other programming languages, where coding practices, comment- and review styles may differ. 
Finally, the rigorous data cleaning employed by some approaches can negatively impact the generalizability.
This is especially true if preprocessing choices, for example, filtering certain types of comment or rebalancing \emph{ChQual} data, create distributions that no longer reflect the real world.

The availability of earlier datasets and other artifacts such as source code and model weights is invaluable for any research.
In our survey, we were able to locate research artifacts published alongside the articles of 16 of the 24 publications surveyed, which confirms a general positive trend in artifact publishing~\cite{Heumueller2020}.
In addition, the size of the available, labeled datasets has grown from ~10k instances~\cite{Tufano2019} to currently 100k-300k instances~\cite{Li2022,Tufano20222291}.
However, despite the obvious benefits, there is still little reuse of data across research groups, as shown in Section \ref{sec-discussion-results}.
Of course, studying new learning tasks or variations often requires more or different kinds of data, so naturally datasets and data collection must evolve.
In this regard, only a few groups (e.g., ~\cite{Heumueller2021a}) have so far published their data collection code, even for common sources like GitHub. 
Thus, we recommend that researchers also share their data collection code.

\finding{External Validity Findings}{Researchers reported that key challenges to external validity concern datasets and data collection.
We specifically advocate for transparency in data preprocessing and for publishing the source code used for data collection and processing alongside the resulting datasets.}

%% file: content/experiment.tex
\section{An Experiment on the Importance of Baselines\label{sec-experiment}}
In this Section we want to illustrate the importance of having appropriate baselines by means of an experiment on the \emph{ChaQual} task.
To us, including appropriate human-understandable baselines is of particular importance for future MCR automation research, since, as explained in Section \ref{sec-discussion-validity}, our survey of past research showed that 15 of 24 publications used only blackblox models as baselines or used no baselines at all.

The upper section of Table \ref{table-experiment}, shows the results of four state-of-science transformer models for \emph{ChQual}, as published by a group of researchers at Microsoft and LinkedIn~\cite{Li20221035}.
As part of their research on pre-training, they fine-tuned their baseline models for the binary change quality estimation task using their publicly available dataset~\cite{Li20221035}.
The baselines in Table \ref{table-experiment} differ primarily in the pre-training tasks used, \emph{Transformer (no pre-training), T5 (pre-training by Tufano et al. ~\cite{Tufano2019})} and \emph{CodeT5 (pre-training by Wang et al. ~\cite{Wang2021})}.
As can be seen, the performance increases in that order, but the table gives no insight into how these performance improvements translate to real-world scenarios.
While 78.6\% precision and 65.65\% recall may at first look like reasonably high numbers for such an ambiguous learning task, it is important to note that the training and validation data was rebalanced by down-sampling instances without comments, which in practice occur 2-3 times as frequently\footnote{Other researchers report even higher degrees of imbalance.}~\cite{Li20221035}.
Thus conclusions regarding the real-world performance are not possible.
As explained in Section \ref{sec-discussion}, we believe that in this kind of situation human-understandable baseline models are a great way to gain a better understanding of the performance of a model.
To explore this idea, we performed an experiment with two simple models using synthetic features.
We used the same balanced dataset (including the original splits) and metrics as the \emph{CodeReviewer} team, so comparing our results to theirs should be valid.
Our implementation is included in the replication package.

\subsection{Experimental Setup}
\paragraph*{Features} The idea was to find a number of simple features that capture the notion of change complexity, as this should likely be correlated with the likelihood of a change being commented.
We came up with a total of 11 features, including the length of the change, old file and new file, the number of additions and deletions, and the proportion of if-tokens and curly-braces compared to the total number of tokens in the change.
The complete list of features is in our replication package.

\paragraph*{Models} As models, we selected a \emph{DecisionTree} and a multilayer perceptron classifier, \emph{MLP}, from scikit-learn~\cite{scikit-learn}.
\emph{DecisionTree} was restricted to 256 leaf nodes.
\emph{MLP} has one hidden layer with 32 units and was trained for 200 iterations.

\paragraph*{Feature Selection} To find the best combination of features, we trained all 2047 possibilities for \emph{DecisionTree} in a multiobjective optimization, attempting to maximize precision and recall, while minimizing the number of features used.
We compared all models in the resulting pareto-front with \emph{CodeReviewer} by computing the sum-of-squared errors for precision and recall.
The closest model uses six features\footnote{Features: patch-length, added-lines, changed-lines, average-token-length, if-rate, old-file-length} and is the one we report here.
To save training time, for \emph{MLP}, we directly used this feature-combination, accepting the possibility that other combinations could have been superior.

\paragraph*{Training} All training was performed on a Lenovo T470p laptop with 64GB of RAM and an Intel(R) Core(TM) i7-7700HQ CPU @ 2.80GHz, without using any GPU.
Training, validating and testing one configuration takes \textapproxnew 10s for \emph{DecisionTree} and \textapproxnew 37s for \emph{MLP}.

\paragraph*{Results}
\begin{table}
	\centering
	\begin{tabular}{lllll}
		\textbf{Model (Layers \#)}    & \textbf{Precision} & \textbf{Recall} & \textbf{F1} & \textbf{Accuracy} \\
		\hline
		Transformer (12)  & 75.50     & 46.07  & 56.93 & 65.16    \\
		T5 (6)            & 70.82     & 57.20  & 63.29 & 66.82    \\
		CodeT5 (12)       & 70.36     & 58.96*  & 64.16 & 67.07    \\
		\textbf{CodeReviewer (12)} & \textbf{78.60}     & \textbf{65.63}  & \textbf{71.53} & \textbf{73.89}    \\
		\hline
		DecisionTree (n.a.) & 74.32 			& 58.84  & 65.68 & 69.26    \\
		MLP (1)             & 75.87*    & 57.93  & 65.70* & 69.75*
	\end{tabular}
	\caption{Experimental Results. Bold typeset marks best- and asterisks mark second-best performing models\label{table-experiment}}
\end{table} The bottom section of Table \ref{table-experiment} shows the metric scores that our two models achieved on the test split of the \emph{CodeReviewer} dataset.
First, they show that neither of the models can quite match the performance of \emph{CodeReviewer}.
However, both models clearly outperform all the other transformer baselines and, of the two, \emph{MLP} achieves a slight edge in precision, bringing it closer to \emph{CodeReviewer}.

\paragraph*{Discussion}
The point of this experiment was not to demonstrate that a simple model can outperform a much more complex transformer model, although with some additional time spent in feature engineering, this may even have been possible.
Instead, we wanted to show that having a simple, understandable baseline provides essential context to better understand the performance of more complex models.
In this case, it also shows that the suitability of the original transformer baselines was questionable.

To the original question ``How good is \emph{CodeReviewer} at estimating code change quality?" we now also have an qualitative answer.
It is only slightly better than a naive model that takes into account six features of code complexity. 
An interesting follow-up experiment would be to combine \emph{CodeReviewer} and the \emph{MLP} model.
If the performance does not increase, it would indicate that the original model had already learned these or equivalent features.
If performance does increase, it would suggest that, given the training data, the transformer was not capable of learning these trivial features.
This, in turn, could inspire new kinds of pre-training tasks, hybrid models, and other improvements.

%% file: content/threats.tex
\section{Threats to Validity\label{sec-threats}}
\paragraph*{Survey Process}
Regarding \emph{RQ1}, our survey may be primarily affected by two threats.
The first is a threat to completeness, since our search query may have missed relevant publications that were identified by neither the selected keywords nor by the snowballing process.
In particular, this is the case for publications that are more recent than the date of the scopus search.
However, since this kind of survey unavoidably takes a lot of time, some cutoff date has to be accepted.
We trust that more recent publications will eventually be covered in future work, either by us or by other researchers.

Further, we could have made mistakes in our filtering and screening process and overlooked or misjudged some relevant publications.
However, especially going backward in time, most publications that we found were cited more than once. 
This redundancy protects against some of these mistakes.
Since we reached saturation after only two iterations, the likelihood of having missed relevant publications before 2015 is low.

The second threat affects our summaries of individual publications.
Even though we spent significant effort analyzing papers, comparing earlier and later work by the same researchers, and in many cases even looking at source code and datasets to clarify details, it is possible that we misunderstood some details.
However, we believe that this is very unlikely to affect the validity of our general contributions.

\paragraph*{Aggregation of Results}
Regarding \emph{RQ2 and 3}, there is the possibility of human mistakes such as copy-and-paste errors or transposed-digits-errors.
Although we did our best to mitigate this threat, it cannot be completely ruled out.
To mitigate this risk, in our evaluation scripts\footnote{Evaluation script is included in our replication package} we have implemented a duplicate value detection across all metrics. 
For exact duplicates, which we consider to be copy-and-paste error candidates, we double checked the values by revisiting the respective publications.
Therefore, we are sure that the remaining duplicate values are coincidental.

\paragraph*{Baselines Experiment}
We used exactly the data, splits, and metrics presented by Li et al. ~\cite{Li2022}, so our \emph{ChQual} experiment could be subject to the same data related risks as theirs.
Also, there is always the risk of bugs in our implementation, even though we checked it multiple times.
For this reason, we publish the source code alongside this article, allowing others to review our implementation and validate the results.
Most importantly, we believe that neither of these aspects would be likely to have a relevant impact on the validity of the general observations and recommendations that we support with this experiment.

%% file: content/relatedwork.tex
\section{Related Work\label{sec-related-work}}
In this section we look at related work which also systematically analyzed the state of science in code review automation focusing on the primary tasks of change quality automation, comment generation, and code refinement.

In their 2024 paper, Tufano et al. analyze the performance of three state-of-the-art models for the comment generation and refinement tasks ~\cite{Tufano2024} . 
They make multiple important contributions (1) analyzing how well these models can predict different types of human changes using a novel taxonomy, (2) reflecting on the suitability of the prevalent code review datasets and (3) report on a first comparison  of specific models with the general purpose LLM ChatGPT.
Compared to their work, our survey is different in focus and much more comprehensive in terms of tasks, research questions, and the number and type of approaches surveyed.

A recent survey by Yang et al. comprehensively covers a broad spectrum of topics, including the evolution of research trends, venues, and a strong empirical and macro understanding of the field ~\cite{Yang2024}. 
Although this survey has a wide-ranging scope, it only briefly touches on code review automation, making it a smaller aspect of the overall study.
In contrast, our research is focused exclusively on code review automation approaches, addresses different research questions, and aims to provide a more granular and detailed examination of MCR automation, particularly by formalizing and zooming in on tasks summarized as "review comment recommendation" and "other techniques" by Yang et al.

In their comprehensive work on \emph{Machine Learning for Big Code and Naturalness} in 2021, Allamanis et al. also survey many different source code related machine learning approaches~\cite{Allamanis2018}. 
However, their work contains only one reference to a code review automation approach.
Further, they identify many important challenges, like finding the right metrics and baseline models, just as we do.  

In their 2018 survey Xiaomeng et al. give a brief overview of machine learning-based vulnerability detection tools, framing the vulnerability detection use case as one aspect of code review ~\cite{Xiaomeng201956}.
Due to this very narrow interpretation, beyond the paper title, there is almost no overlap with our work.

The challenge of interpreting results, including the question of appropriate baselines, is a general challenge in deep learning and has been studied in other fields of deep learning~\cite{Mukhoti2018}. 
However, since our work is focused on code review automation, our work is more specific with regard to challenges and recommendations.

In their 2021 paper, Hellendoorn et al.~\cite{Hellendoorn2021} also discussed some of the challenges of automating code review in real-word scenarios.
This publication is included in our survey because of its experimental work and their brief discussion on the challenges and research directions is part of what inspired us to do this study.

Related but outside this scope, Cetin et al. published a comprehensive survey on reviewer recommendation~\cite{Cetin2021}.

%% file: content/conclusion.tex
\section{Conclusion\label{sec-conclusion}}
In this work, we have made several contributions aimed at supporting researchers in the field of code review automation. 
First, we formalized the primary code review tasks of change quality estimation, comment generation, and code refinement. 
Using these definitions, we systematically identified and surveyed 24 code review automation approaches from a set of 691 candidate publications spanning 2015 to April 2024. 
We categorized them and highlighted their main contributions to provide a comprehensive view of the evolution and current state of code review automation research.

Through four research questions, we analyzed existing approaches, examined the use of evaluation metrics and how to apply them effectively, assessed the current state of research for each learning task, and highlighted key validity challenges along with strategies to mitigate them. 
To further emphasize the importance of human-understandable baselines as reference points for interpreting results, we conducted an experiment with two novel and reusable synthetic-feature baseline models for change quality estimation.

We hope that our findings and contributions will not only support individual researchers in navigating and advancing code review automation but also contribute to the broader progress of the field by contributing to the effort for more standardized methodologies, data, and reproducible research.